%% file: main-v13-arxiv.tex
\documentclass[prx,twocolumn,superscriptaddress,nofootinbib,longbibliography,aps]{revtex4-2}
\usepackage{dsfont} %
    \usepackage[rgb]{xcolor}
\usepackage{makecell}

\usepackage{bm} %

\usepackage[normalem]{ulem}
\usepackage{bbold} %
\usepackage{times}
\usepackage{graphicx,braket}
\usepackage{hyperref}
\hypersetup{colorlinks=true,linkcolor=blue,citecolor=blue}
\usepackage{amsmath,amssymb,amsfonts}
\usepackage{wrapfig}

\usepackage{siunitx}

\usepackage{cleveref}%
\crefname{equation}{Eqs.}{Eqs.}
\Crefname{equation}{Equation}{Equations}%
\crefrangelabelformat{equation}{(#3#1#4--#5#2#6)}
\crefmultiformat{equation}{Eqs. (#2#1#3}{, #2#1#3)}{#2#1#3}{#2#1#3}
\Crefmultiformat{equation}{Equations (#2#1#3}{, #2#1#3)}{#2#1#3}{#2#1#3}

\usepackage{booktabs}
\usepackage{mathtools}

\newcommand{\ketbra}[2]{\left| #1 \rangle \langle #2 \right|}

\newcommand{\dB}[1]{\SI{#1}{\decibel}}

\newcommand{\abs}[1]{\left|#1\right|}

\newcommand{\im}{\text{Im}}

\newcommand{\figref}[1]{\mbox{Fig.~\ref{#1}}}

\newcommand{\appref}[1]{\mbox{SI~\ref{#1}}}
\renewcommand{\eqref}[1]{\mbox{Eq.~(\ref{#1})}}

\newcommand{\be}{\begin{equation}}
\newcommand{\ee}{\end{equation}}
\newcommand{\bea}{\begin{eqnarray}}
\newcommand{\eea}{\end{eqnarray}}

\usepackage{outlines}

\begin{document}

\flushbottom
\title{Squeezed Cooling and Lasing with Cold Atoms in Cavity-QED}

\author{Rodrigo Grande de Diego}
    \affiliation{Quantum Advanced Research Center (QuARC), CSIC, Calle Serrano 113b, 28006 Madrid, Spain}
	\affiliation{
Institute of Fundamental Physics (IFF), CSIC, Calle Serrano 113b, 28006 Madrid, Spain}

\author{Carlos S\'anchez Mu\~noz}
    \email{c.sanchez.munoz@csic.es}
    \affiliation{Quantum Advanced Research Center (QuARC), CSIC, Calle Serrano 113b, 28006 Madrid, Spain}
	\affiliation{
Institute of Fundamental Physics (IFF), CSIC, Calle Serrano 113b, 28006 Madrid, Spain}

\begin{abstract}
A squeezed laser undergoes a lasing transition into a squeezed cavity mode, where stimulated emission generates a bright coherent field with a laser-like coherence time while maintaining one field quadrature below the vacuum noise level.
Here we propose an optical implementation
that needs neither a nonlinear crystal nor injected squeezed light, based
on engineered interactions between ensembles of rubidium atoms and a squeezed cavity mode via cavity-assisted Raman transitions.
A first ensemble cools the cavity into a squeezed vacuum, a mechanism we term squeezed cooling, yielding up to \dB{9} in the intracavity field, surpassing the \dB{3} limit of optical parametric oscillators.
Tuned for output instead, it reaches up to 18 dB of output squeezing.
Incoherently pumping a second ensemble drives a lasing transition into the squeezed mode,  emitting a bright beam with long coherence time and noise below vacuum spanning a megahertz-wide band, away from the sharp phase-diffusion peak of excess noise at the carrier frequency, and robust to  relaxation oscillations. We characterize lasing thresholds, superradiant emission, and linewidth narrowing, and identify the experimentally accessible regime where the system exhibits the defining features of a squeezed laser.
\end{abstract}
\date{\today} \maketitle

Squeezed light with quadrature noise below the vacuum limit is a central resource in quantum technologies, enabling key applications in fields like quantum metrology \cite{aasi_enhanced_2013, schnabel2017a, lawrie_quantum_2019},  cryptography \cite{marino_deterministic_2006,wang_continuous-variable_2019,derkach_squeezing-enhanced_2020}, and information processing \cite{braunstein_quantum_2005}.
Of particular interest are \emph{bright squeezed states} \cite{paschotta_bright_1994, schmitt_photon-number_1998, pontula2025}, which combine macroscopic coherent amplitude with reduced fluctuations, making them attractive for optical interferometry~\cite{schnabel2010,aasi_enhanced_2013,schnabel2017a,demkowicz-dobrzanski2015a}, coherent light-matter interfaces~\cite{casacio2021, dorfman2016}, or any other application that may benefit from both strong signal and reduced noise~\cite{paschotta_bright_1994,chekhova2015}.

Squeezing in the optical regime is routinely produced from the output field of a sub-threshold degenerate optical parametric oscillator (OPO)~\cite{gardiner1984, wu1987, vahlbruch2016}.
While this platform is well established for producing a squeezed output field, it does not simultaneously provide high \emph{intracavity} squeezing, bright emission, and a coherent carrier of its own within a single setting. In particular, the intracavity noise reduction is fundamentally limited to half of the vacuum noise ($\approx\ $\dB{3})~\cite{milburn_production_1981, qin_beating_2022} due to single-photon losses. This limitation hampers applications based on intracavity squeezing rather than on squeezed traveling waves, such as enhanced light–matter interactions~\cite{burd2021}, qubit readout~\cite{qin_beating_2022, didier2014}, position sensing~\cite{peano2015}, and the generation of nonclassical states~\cite{groszkowski2020,chen2021}.
For applications requiring narrow linewidth and phase coherence, such as resonant excitation of narrow atomic transitions~\cite{wolfgramm2013a} or quantum spectroscopy~\cite{dorfman2016,casacio2021},  OPO sources cannot operate above threshold to benefit from a carrier of their own without squeezing collapsing into a classical mixture of coherent states~\cite{drummond1981,wolinsky1988,mason1998,leghtas2015}. Displacing the sub-threshold output with an external coherent field yields a bright squeezed beam~\cite{oelker2014,li2025}, but its carrier is borrowed from a separate laser, to which the squeezing must be phase-locked.

All these limitations have motivated the development of a new class of squeezed light source: the \emph{squeezed laser}~\cite{navarrete-benlloch_inducing_2014, sanchez_munoz_squeezed_2021, tian2025, zhang2026,lee_bath-free_2026}. These systems are engineered to undergo a lasing phase transition in a squeezed cavity mode, offering a compelling platform for the generation of bright squeezed states, with the added benefit of the laser-like coherence time, potentially enabling the use of squeezing for applications where this property is crucial, such as the detection of gravitational waves~\cite{Kwee2012}, definition of fundamental quantities~\cite{ludlow2008, jiang2011}, or sideband cooling~\cite{teufel2011}.

    \begin{figure*}[t!]
        \centering
        \includegraphics[width=1\textwidth]{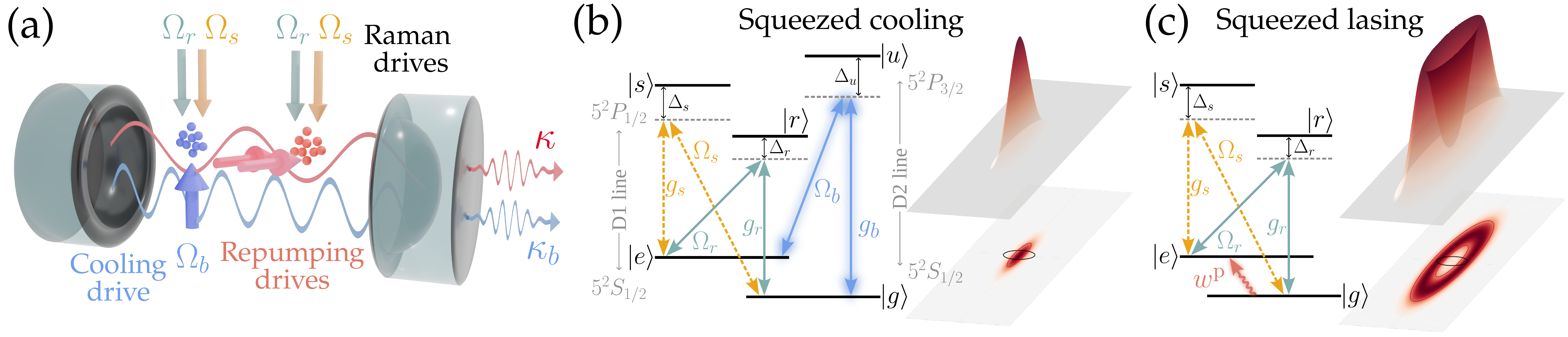}
        \caption{\textbf{Implementation of the squeezed laser}. (a) Scheme of the proposed physical setup of the system. The two ensembles of rubidium atoms are trapped within an optical cavity and transversely pumped by external lasers. The blue colored ensemble contains atoms from Fig. 1(b), and the red one contains atoms from Fig. 1(c). (b) Energy level diagram of the cooling atoms. In the long-time limit, these act as a buffer, cooling the cavity to a squeezed vacuum state, represented with its Wigner function. (c) Energy level diagram of the lasing atoms that constitute the gain medium. When these are pumped with the repumping drives, the steady state can be accurately described by a phase-diffused mixture of displaced squeezed states. The black circle in both plots of the Wigner functions represents vacuum fluctuations.}
        \label{fig_1}
    \end{figure*}

Optical squeezed lasers where both squeezing and bright signal are autonomously built in within the system without any injection of external squeezing, as required in current optical schemes~\cite{sanchez_munoz_squeezed_2021, tian2025}, have not yet been proposed.
 Here, we present one by introducing an atomic cavity-QED scheme designed to meet the two requirements for a squeezed laser. The first is an engineered coupling between the atoms and a squeezed cavity mode, and the second is an effective interaction between the cavity and a squeezed reservoir, equivalent to standard single-photon loss in the squeezed-mode basis, driving the system into a squeezed vacuum~\cite{cirac1993, kienzler2015,kronwald2013,wollman2015,dassonneville2021}. The engineered dissipation, termed here \emph{squeezed cooling}, allows one to stabilize a squeezed vacuum beyond the conventional \dB{3} limit, and, when adding incoherent pumping of the active medium, induces a squeezed lasing transition.
 For realistic cold-atom parameters, the squeezed vacuum reaches reaches \dB{9} of intracavity squeezing, and the laser emits a bright, phase-coherent beam with noise below vacuum level over a megahertz-wide band, robust against its relaxation oscillations.

 The scheme uses the hardware of superradiant Raman lasers~\cite{meiser2009,bohnet2012,norcia_cold-strontium_2016}, based on the realization of anisotropic Dicke models \cite{baumann_dicke_2010, baumann_exploring_2011, brennecke_real-time_2013, klinder_dynamical_2015,dareau_observation_2018} via cavity-assisted Raman transitions~\cite{dimer_proposed_2007},
 thus relying on a mature technique~\cite{baden_realization_2014,zhiqiang_nonequilibrium_2017,zhang_dicke-model_2018} and
 adding control over quantum noise to these optical references. Here, squeezed cooling and lasing are respectively engineered by two separate atomic ensembles of the same species (see Fig.~\ref{fig_1}), coupled to the main lasing cavity mode via two shared Raman drives.

\section*{Results}
\subsection*{Implementation}
 We consider a total of $N$ $^{87}$Rb atoms trapped within an optical cavity. These atoms are spatially distributed into two different ensembles within the cavity, so that highly focused laser beams can act individually on each ensemble [see \figref{fig_1}(a)]. We refer to them as \emph{cooling} and \emph{gain} ensembles, based on the role they play. The optical cavity is assumed to have a high quality factor in one of its modes, of frequency $\omega_\text{cav}$, which we use as lasing mode and describe by the annihilation operator $\hat{a}$ and decay rate $\kappa$. To do reservoir engineering, we also make use of another cavity resonance $\hat b$, assumed to have a higher frequency $\omega_b$ and a much larger decay rate $\kappa_b$, playing the role of a buffer mode.
The atoms are described in terms of hyperfine Zeeman sublevels, whose degeneracy is broken by an external magnetic field. We assume that the atomic dynamics are restricted to a four-level manifold (see Methods) comprising two ground states $\ket{g}\equiv\ket{F=1,\ m_F=-1}$ and $\ket{e}\equiv\ket{F=2,\ m_F=-2}$, and the two  $5^2 P_{1/2}$ excited states of the D1 line
$\ket{r}\equiv \ket{F'=1,\ m_{F'}=-1}$ and $\ket{s} \equiv \ket{F'=2,\ m_{F'}=-2}$, as well as an extra excited state of the D2 line, $|u\rangle \equiv\ket{F'=1,\ m_{F'}=-1} $, in the cooling ensemble.  The ground states $\{\ket{g},\ket{e}\}$ of the $i$-th atom define an effective two-level system (TLS) with lowering operator $\hat\sigma_i \equiv \ketbra{g}{e}_i$.

Coupling of all atoms to a squeezed cavity mode is enabled by two transverse Raman lasers on the D1 line with Rabi frequencies  $\Omega_s$ and $\Omega_r$, driving transition $|g\rangle \leftrightarrow |s\rangle$ and  $|e\rangle \leftrightarrow |r\rangle$ respectively [see Fig.~\ref{fig_1}(b,c)]. Their frequencies, $\omega_{Ls}$ and $\omega_{Lr}$, are tuned to the two-photon resonance~\cite{dimer_proposed_2007} that yields a Hamiltonian interaction term $\lambda_r \hat \sigma_i \hat a^\dagger+ \lambda_{cr} \hat \sigma_i \hat a + \text{H.c.}$ , where $\lambda_r \equiv -g_r\Omega_r/2\Delta_r$ and $\lambda_{cr}\equiv -g_s\Omega_s/2\Delta_s$ ($g_{r,s}$ being cavity coupling rates and $\Delta_{r,s}$ the Raman detunings, see Methods).

Fixing specific values for the optically-tunable parameters $\omega_{Ls}$, $\omega_{Lr}$, and $\Omega_r$, to remove other unwanted residual terms (see Methods), one obtains, in a frame rotating at the reference frequency $\omega_0=(\omega_{Ls}+\omega_{Lr})/2$ for cavity $\hat a$, an effective Jaynes-Cummings coupling between the TLSs and a squeezed cavity mode,
\begin{equation}
    \hat H_{R,\text{eff}} = \sum_{j=1}^N g_0(\hat a_s^\dagger \hat\sigma_j + \text{H.c.}),
    \label{eq:H_JC}
\end{equation}
where the squeezed mode is defined in terms of a Bogoliubov transformation as $ \hat a_s \equiv \hat a \cosh r  + \hat a^\dagger\sinh r$, with the squeezing factor $r \equiv  \text{atanh}[\lambda_{cr} /\lambda_r]$, and the coupling rate given by
\begin{equation}
    g_0= \sqrt{\lambda_r^2-\lambda_{cr}^2}.
    \label{eq:g_0}
\end{equation}

The resulting effective, resonant Jaynes-Cummings coupling between the squeezed cavity   and both the cooling and the gain ensembles is the foundation for the mechanisms of squeezed cooling and squeezed lasing that we describe in the next sections.

 Cooling is enabled by a third laser with Rabi frequency $\Omega_b$ driving the transition $|e\rangle \leftrightarrow |u\rangle$ in the cooling ensemble, in resonance with the cavity-assisted Raman process that couples it to the lossy mode  $\hat b$ [see Fig.~\ref{fig_1}(b) and Methods].  Finally, squeezed lasing can be realized if the gain atomic ensemble is incoherently pumped via weak repumping lasers (see Methods). A joint description of the Raman coupling of the cooling ensemble to the buffer mode $\hat b$, the incoherent pumping and the Raman-induced local dissipation, yields the full model $\mathcal L_\text{full}$ of \eqref{eq:me_adiabatic_full} (see Methods), which only assumes the elimination of the atomic excited states and against which we benchmark all effective descriptions below. Parameters are listed in Table~\ref{tab:parameters}.

\subsection*{Squeezed cooling}
\begin{figure*}
\centering
\includegraphics[width=1\linewidth]{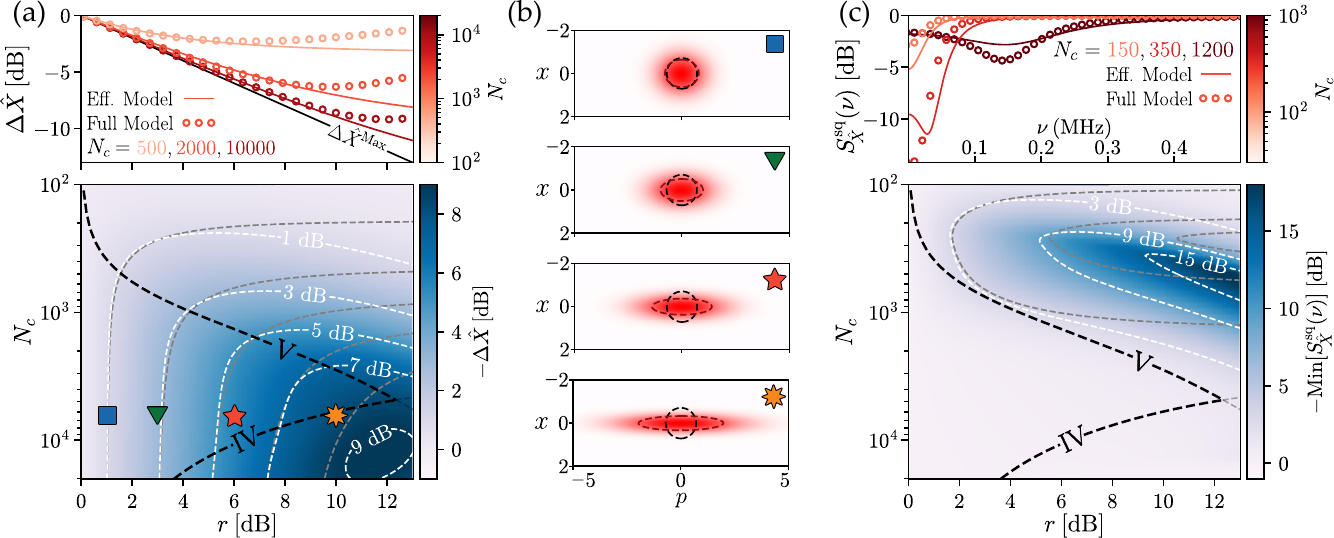}
\caption{\textbf{Squeezed Cooling Mechanism}. (a) Numerical results (via the cumulant method) for the intracavity squeezing in the steady state of \eqref{eq:me_eff_full_cooling}. We plot the quadrature-noise reduction in decibels, computed as $-20\log_{10}({\Delta\hat X})$. Black dashed curves indicate where the different relations from Table~\ref{tab:conditions} become equalities. The area within the bold segments corresponds to the parameter space where inequalities hold. The white contour lines of the color map can be compared with the gray dashed lines, which represent the results obtained from the effective model in \eqref{eq:me_sq_cool}. In the upper panel we plot the results from both models for several values of $N_c$, together with the reference for the maximum possible squeezing achievable. (b) Wigner functions of the cavity steady state of \eqref{eq:me_HP}---a version of~\eqref{eq:me_eff_full_cooling} with bosonized spins---for four different values of $r$ at $N_c=6500$, with the black circles marking vacuum fluctuations. These were computed via exact master-equation simulations using \texttt{QuantumToolbox.jl} \cite{mercurio2025}. (c) Numerical results (via the cumulant method) for the squeezing spectrum in the steady state of \eqref{eq:me_eff_full_cooling}. We plot the value of the spectrum in decibels at the sideband frequency that yields the maximum output squeezing, computed as $-10\log_{10}({\text{Min}[S_{\hat{X}}^\text{sq}(\nu)]})$. Curves in the colormap are analogous to those of panel (a). Above we show the spectrum profile from both full and effective models at $r=1$ for several values of $N_c$.}
\label{fig_2}
\label{fig_2b}%
\end{figure*}
The role of the $N_c$ atoms from the cooling ensemble is to serve as a squeezed buffer for the cavity mode $\hat a$, resulting in an effective decay term $\mathcal D[\hat a_s]$  that cools the cavity into a squeezed vacuum. Such behaviour is enabled by the Raman coupling between the cooling atoms and the squeezed cavity mode in \eqref{eq:H_JC}, and the  Raman coupling to the buffer cavity mode $\hat b$ [\eqref{eq:H_cool_eff} in Methods].
Adiabatically eliminating first the buffer mode and then the cooling ensemble (see Methods), we obtain the reduced master equation
\begin{equation}
    \mathcal L_\text{cool,eff}\,\hat{\rho} = -i\left[\Delta\hat{a}_s^\dagger\hat{a}_s,\hat{\rho}\right]+\frac{\kappa}{2}\mathcal{D}[\hat{a}]+\frac{\tilde{\kappa}}{2}\mathcal{D}[\hat{a}_s],
    \label{eq:me_sq_cool}
\end{equation}
with
\begin{equation}
    \Delta\approx-\frac{4\tilde{\delta}_0g_0^2}{N_c\Gamma^2},\quad\tilde\kappa \approx \left(\frac{g_0}{g}\right)^2\kappa_b, \label{eq:sq_kappa_simp}
\end{equation}
where $\tilde\delta_0$ is a small Lamb shift (defined in Methods) and where we have already assumed that the number of cooling atoms is large enough to fulfill $ N_c \Gamma\gg |\tilde\delta_0|$. In this limit of large $N_c$ we can neglect $\Delta$ relative to $\tilde\kappa$, yielding a simple expression for the intracavity   fluctuations of the squeezed quadrature $\hat{X}\equiv \hat{a}+\hat{a}^\dagger$ in the steady state of \eqref{eq:me_sq_cool},
\begin{equation}
\Delta\hat{X}=\sqrt{\frac{e^{-2r}+\kappa/\tilde\kappa}{1+{\kappa}/{\tilde{\kappa}}}},
    \label{eq:DeltaX2-simplified}
\end{equation}
with $r$ being the target squeezing parameter
set via the Raman drives as $\tanh r=\Omega_{s}/2\Omega_{r}$. The full expression for finite $N_c$ is in Supplementary Information (SI)~\ref{app:eliminations_cooling}.

According to \eqref{eq:DeltaX2-simplified}, the effective degree of squeezing is established by the ratio ${\kappa}/{\tilde{\kappa}}$, reaching a perfect agreement with the squeezed target value, $\Delta\hat{X} \approx e^{-r}$,  in the limit
${\tilde{\kappa}}/{\kappa}\gg\left(e^{2r}-1\right)$ (condition V in Table~\ref{tab:conditions}). Note that one can choose any squeezing angle by tuning the complex phases of Rabi frequencies, which we assumed here to be real.

We confirm these conclusions by numerical simulations of the steady state of the full system without adiabatic eliminations [$\mathcal L_\text{cool,full}$ in \eqref{eq:me_eff_full_cooling}] via the cumulant expansion approach \cite{kubo1962}, fixing a set of parameters, listed in  Table~\ref{tab:parameters}, consistent with typical values in atomic cavity-QED experiments \cite{arnold_self-organization_2012,baden_realization_2014,zhiqiang_nonequilibrium_2017,steckRb87Dline,zhang_dicke-model_2018}. The results are shown in  \figref{fig_2}(a,b), depicting the intracavity steady-state quadrature-noise reduction $\Delta\hat X$ as a function of $N_c$ and $r$. In such parameter space, the region of best agreement with the target squeezing matches the region of validity of the adiabatic elimination that leads to $\mathcal L_\text{cool,eff}$, (see Table~\ref{tab:conditions}),
which sets the minimum number
of cooling atoms required for a given target squeezing (condition V) and an upper limit to the squeezing of $\sim\,$\dB{9} (condition IV)  for our choice of realistic parameters. This value clearly outperforms the maximum intracavity squeezing achievable in other widespread platforms such as OPOs, where this magnitude is limited to $\approx\,$\dB{3} due to single-photon loss~\cite{milburn_production_1981, qin_beating_2022}.

The squeezing in the output of $\hat a$ shows a very different behavior from the intracavity squeezing. We note that \eqref{eq:me_sq_cool} describes a cavity that, in the squeezed basis, behaves as a linear two-port with two outputs (physical mirror of $\hat a$, set by $\kappa$, and cooling bath, set by $\tilde \kappa $). Thus, according to standard input-output theory, in the limit $\Delta\ll (\kappa, \tilde\kappa)$, the output squeezing of the $X$ quadrature $S_{\hat X}^\text{sq}(\nu)$ (defined at sideband frequencies  $\nu$ around the reference frequency $\omega_0$ set by the Raman drives) is maximized at $\nu=0$ when $\tilde\kappa=\kappa$, contrasting with the regime ${\tilde{\kappa}}/{\kappa}\gg\left(e^{2r}-1\right)$ that optimizes intracavity squeezing (see SI \ref{app:emission_spec}). A finite value of $\Delta$ modifies the optimum condition for $\tilde\kappa$ and allows to further improve the output squeezing, making the system act as a below-threshold parametric amplifier and surpass the target squeezing given by $r$: in the bare basis, $\Delta\hat a_s^\dagger\hat a_s$ contains the two-photon term $\frac{\Delta}{2}\sinh 2r\,(\hat a^2+\hat a^{\dagger 2})$, a crystal-free parametric drive.

This claim can be checked by scanning the same parameter space as in \figref{fig_2}(a) and registering the minimum value of the squeezing spectrum of $\mathcal L_\text{cool,full}$. The results obtained are shown in \figref{fig_2}(c), and they confirm that the squeezed cooling mechanism operates in two different regimes depending on whether condition V is fulfilled or not. In the first regime, the engineered dissipation $\tilde \kappa$ dominates, yielding an excellent intracavity squeezing that, however, is not translated into the output. In the second regime, both $\kappa$ and $\tilde \kappa$ input ports are balanced in such a way that, despite obtaining a mediocre intracavity squeezing, the output state presents a very high squeezing, which, due to the non-negligible values of $\Delta$, displays noise reductions of up to $\sim$\dB{18}, above the target set by $r$. Since $|\Delta|/\tilde\kappa\propto 1/N_c$, the number of cooling atoms tunes the device from a dissipative squeezed reservoir (best intracavity squeezing) to a parametric amplifier (best output squeezing).

\subsection*{Squeezed lasing}
Having established the squeezed cooling approach, we now consider autonomous squeezed lasing action, achieved when  both the cooling and gain ensembles are active and operating within the region of parameters that we have established (see I-IV from Table~\ref{tab:conditions}). In that case,
 the complete effective description of the system of gain atoms plus lasing cavity will be given by the master equation
   \begin{multline}
        \mathcal L_\text{las}\,\hat\rho = -i\left[\Delta\hat{a}_s^\dagger\hat{a}_s,\hat{\rho}\right]+\frac{\kappa}{2}\mathcal{D}[\hat{a}]+\frac{\tilde{\kappa}}{2}\mathcal{D}[\hat{a}_s]+\sum_{j=1}^{N_g}
        \left\{\frac{w}{2}\mathcal{D}[\hat{\sigma}_j^\dagger] \right.
        \\
        \left.
                +\frac{\gamma}{2}\mathcal{D}[\hat{\sigma}_j]
        +\frac{\gamma_\phi}{2}\mathcal{D}[\hat\sigma_{z,j}]
        +
        g_0\left[\hat a_s^\dagger \hat{\sigma}_j + \hat a_s\hat \sigma_j^\dagger,\hat{\rho}\right]
        \right\}
       ,
    \label{eq:me_lasing}
    \end{multline}
where $w$, $\gamma$ and $\gamma_\phi$ denote the total incoherent pumping, decay and pure dephasing rates, respectively. These include both the direct pumping induced by the repumping mechanism and the rates set indirectly by Raman drives: $w \equiv w^\text p + w^\text{R,a}$, $\gamma \equiv\
\gamma^\text{R,a}$ and $\gamma_\phi = \gamma_\phi^\text{R,a}$.

\eqref{eq:me_lasing} corresponds to the master equation of a standard laser operating in the squeezed cavity mode $\hat a_s$, except for an extra single-photon decay channel for mode $\hat{a}$, detrimental for the intracavity squeezing, since it introduces unwanted noise into the squeezed basis, an effect that becomes negligible when condition V is fulfilled, i.e. $\tilde\kappa\gg\kappa(e^{2r}-1)$. As seen above, this is not a limitation when we are interested in optimizing the squeezing of the output field.

\subsubsection*{Lasing thresholds and stability}
\begin{figure*}
    \centering
    \includegraphics[width=\textwidth]{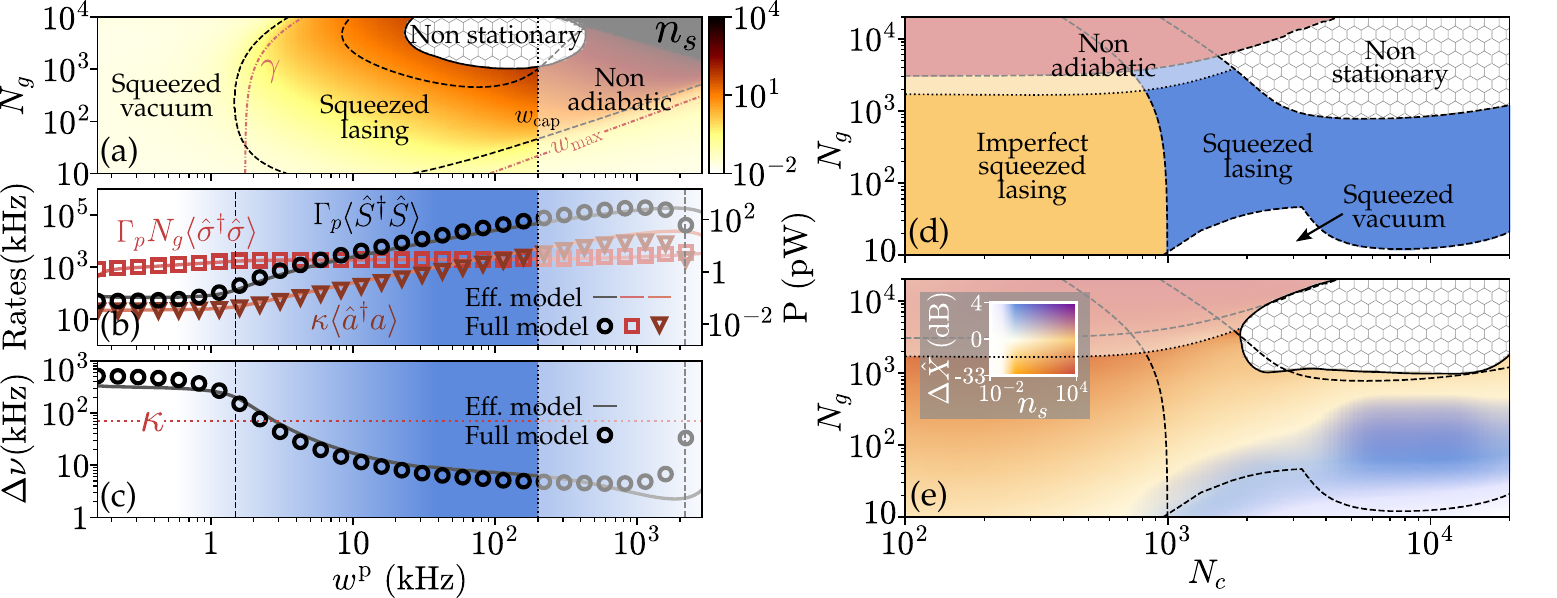}
    \caption{\textbf{Squeezed lasing phase diagram}.
        (a) Stationary photonic population of the squeezed cavity mode in the $(w^\text p,N_g)$ space for $N_c=3000$ [obtained by solving \eqref{eq:me_adiabatic_full} via cumulant expansion].
        Red dash-dotted lines mark the simplified minimum and maximum lasing thresholds in \eqref{eq:wmin-wmax}.
        Dashed black lines mark the mean-field thresholds for lasing [full model, \eqref{eq:me_adiabatic_full}] and for non-stationary dynamics [effective model, \eqref{eq:me_lasing}]. The honeycomb pattern marks the region where the cumulant expansion of \eqref{eq:me_adiabatic_full} yields non-stationary solutions. (b) Emission rate of the lasing ensemble (black circles) at a fixed value of $N_g=400$, together with emission rate of an uncorrelated ensemble (red squares), and photon flux (brown triangles), which is proportional to the lasing power (in the right axis). (c) Linewidth of the emission spectrum at $N_g=400$. In (b,c), results from \eqref{eq:me_adiabatic_full} (markers) are compared with predictions from \eqref{eq:me_lasing} (solid lines). (d) Phase diagram of the $(N_c,N_g)$ space for $w^\text{p}=\min(w_\text{pk},w_\text{cap})$. The boundary between imperfect and genuine squeezed lasing indicates where condition V from Table~\ref{tab:conditions} becomes an equality. The red area corresponds to the region where photonic population becomes too big for the cooling TLSs to remain in the ground state (see SI \ref{app:failure}). (e) Bivariate colormap showing the intracavity photon population and the noise reduction of the squeezed quadrature after breaking phase invariance with an external field. Computed via second-order cumulant equations of \eqref{eq:me_adiabatic_full}. All results fix $r=0.5$, and the white shadowed areas indicate that parameters in that region violate any of the conditions required for the adiabatic elimination of the excited atomic levels (I from Table~\ref{tab:conditions}).
    }
    \label{fig_3}
\end{figure*}

A mean-field ansatz for the effective master equation (\ref{eq:me_lasing}) yields a lower and upper threshold for lasing, as already reported for Raman lasers~\cite{meiser2009,meiser2010,bohnet2012,tieri2017,debnath_lasing_2018}. Neglecting $\Delta$, these are approximately given by the following minimum and maximum pumping rates
\begin{equation}
     w_\text{min} \approx \gamma\quad\text{and}\quad w_\text{max} \approx N_g\Gamma_p,
     \label{eq:wmin-wmax}
\end{equation}
where we defined the single-atom Purcell rate,
\begin{equation}
    \Gamma_p=\frac{4g_0^2}{\kappa+\tilde\kappa},
    \label{eq:purcell-p}
\end{equation}
and where we have assumed $\gamma,\ \gamma_\phi \ll N_g \Gamma_p$. These predicted thresholds are shown as red dashed-dotted lines in Fig.~\ref{fig_3}(a), which represents a phase diagram in the $(w^\text p,N_g)$ space. The complete mean-field stability analysis of \eqref{eq:me_lasing} can be found in SI~\ref{app:stability_analysis}, where we derive the general threshold condition for lasing.
Within the lasing region, the pump rate maximizing the intracavity photon number predicted by the effective model is given by $w_\text{pk}\approx N_g \Gamma_p/2$, with peak photon number $ n_{s,\text{pk}} \approx \frac{1}{8}\frac{\Gamma_p N_g^2}{(\kappa + \tilde \kappa)}$. For most values of $N_g$, however, $w_\text{pk}$ lies beyond the cap $w_\text{cap}= 200\,\text{kHz}$ required by the adiabatic elimination of excited levels (see Methods), which sets instead the maximum achievable lasing power.
Our exploration of $w^\text{p}$ in Fig.~\ref{fig_3} extends beyond $w_\text{cap}$, into a region that we flag as ``non-adiabatic'', to account for the possibility of other incoherent-pumping schemes that may surpass this limit.

More precise phase-space boundaries can be established with a numerical mean-field analysis of the complete model $\mathcal L_\text{full}$ in \eqref{eq:me_adiabatic_full}, which retains all atomic ensembles and cavity modes, and only assumes elimination of atomic excited states.
The resulting threshold is drawn as black-dashed lines in Fig.~\ref{fig_3}(a), showing a qualitative agreement with the boundaries from the effective model in \eqref{eq:wmin-wmax}, but predicting an upper threshold at lower values of $w^\text p$.
The lasing phase also contains a region of non-stationary dynamics, with the
system evolving toward limit cycles or even strange attractors
in the long-time limit [dashed black line and honeycomb pattern in Fig.~\ref{fig_3}(a); see Methods].
We test all these mean-field predictions against a more accurate numerical treatment of the full model $\mathcal L_\text{full}$ (\ref{eq:me_adiabatic_full}), again based on a cumulant expansion truncated at second-order correlators, which captures quantum fluctuations and finite-size effects.

The heatmap in Fig.~\ref{fig_3}(a) represents the intracavity photon number in the squeezed basis, $n_s\equiv\langle \hat a_s^\dagger\hat a_s\rangle$, with regions of high photon population which are confined within the lasing boundaries predicted by mean field, now turned into smooth crossovers. Here, a population exceeding one stationary squeezed photon at $w_\text{cap}$ is reached for $N_g\gtrsim 40$. This lasing phase exhibits a superradiant nature, evidenced by the atomic emission rate exceeding its counterpart for uncorrelated atoms, $\langle \hat S^\dagger \hat S\rangle > N_g \langle \hat\sigma^\dagger \hat\sigma\rangle$, as shown in Fig.~\ref{fig_3}(b). This result is expected for Raman laser setups which are deep in the bad-cavity limit, and have been exploited for superradiant lasing, where coherence is stored in the collective atomic degree of freedom rather than in the intracavity field~\cite{vrijsen2011,meiser2009,bohnet2012a, bohnet2012,meiser2010,tieri2017,debnath_lasing_2018,norcia_cold-strontium_2016,zhang_ultranarrow_2021,zhang_extremely_2024,yu2024}.
The transition to lasing is also signaled by the pronounced linewidth narrowing  of the emission spectrum, $\Delta\nu \ll \kappa$ [see Fig.~\ref{fig_3}(c)]. This narrowing can be related to the closing of the Liouvillian gap~\cite{sanchezmunoz2019} and, consequently, to the system's asymptotic relaxation time, reflecting the longer phase-diffusion timescale characteristic of coherent laser emission (see \appref{app:emission_spec} for details on the calculation of the Liouvillian gap and the emission spectrum).
The sweep in $N_g$ represents a smooth crossover~\cite{tieri2017,norcia_cold-strontium_2016} from steady-state squeezed superradiance---akin to proposals such as Ref.~\cite{dalla_torre_dissipative_2013}---to squeezed lasing.

Overall, these results show that, in the optimal operating regime for output power $N_g \in (10^2,10^3)$, high enough to lase but low enough to avoid the non-stationary regime, the laser sits in the crossover where coherence is shared between the light and the atoms~\cite{debnath_lasing_2018,norcia_cold-strontium_2016}. This operating point yields an output power that can reach tens of pW [see Fig.~\ref{fig_3}(b)], however, note that we operate under a set of parameters chosen to optimize intracavity squeezing, not output power.

\subsubsection*{Intracavity squeezing and quantum fluctuations}

\begin{figure*}
    \centering
    \includegraphics[width=\textwidth]{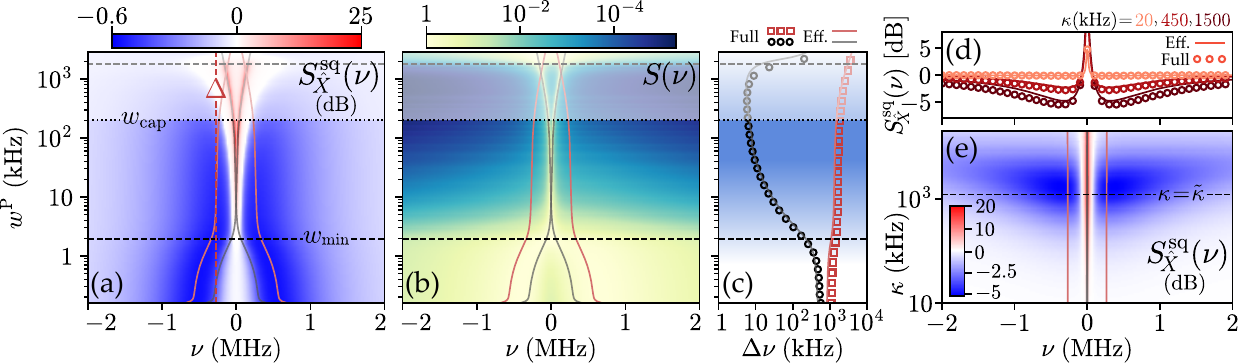}
    \caption{\textbf{Squeezed lasing spectral properties}.
    (a) Steady state squeezing spectrum and (b) emission spectrum of \eqref{eq:me_adiabatic_full} as a function of the pumping rate $w^\text{p}$ for $N_c=5000$, $N_g=300$ and $r=0.8$. $S_{\hat X}^\text{sq}(\nu)$ is expressed in decibels, and $S(\nu)$ is normalized so that its maximum is 1 for every value of $w^\text{p}$. The black dotted line marks $w_\text{cap}=200\ \text{kHz}$, so the shaded area of the plots corresponds to a non-adiabatic regime. The dashed lines indicate the mean-field lasing thresholds. The positions of the carrier peaks obtained for the effective model of \eqref{eq:me_lasing} are marked as grey curves while the red ones correspond to the squeezed sideband dips. (c) Width of the Lorentzian components of the spectra. (d,e) Squeezing spectrum profile for a fixed $w^\text{p}=16\ \text{kHz}$ and different values of $\kappa$.}
    \label{fig4}
\end{figure*}

The features discussed so far are common to Raman lasers of any kind. We now focus on the property that sets our scheme apart, reduced quadrature fluctuations.  First  we consider the squeezing stored in the intracavity field.

The squeezed behaviour is governed by the ratio $\tilde\kappa/\kappa$, which, for a fixed cavity design, depends on the cooling atom number $N_c$. \figref{fig_3}(d) shows a phase diagram in the $(N_c,N_g)$ space, where, at each point, the pump rate is set to the value providing the maximum possible output power without exceeding the cap, i.e., $w =\min(w_\text{pk},w_\text{cap})$.
With this approach, we again find a series of phases. For $N_c$ not high enough, the cooling mechanism is ineffective, as condition V (Table~\ref{tab:conditions}) is not fulfilled. The result is a loss-mismatched, imperfect squeezed lasing phase (ochre), contaminated by the squeezed noise that standard single-photon loss brings into the squeezed basis, as described by \eqref{eq:DeltaX2-simplified}.
This is also evidenced by $\langle \hat a_s^2\rangle$ having appreciable values in the phase-diffused states, meaning these differ from the usual $U(1)$-symmetric lasing states.
We elaborate more on this point in \appref{app:importance_sqc}.

On the other hand, large enough $N_c$ but low $N_g$ yields a region of squeezed vacuum (white), where squeezed cooling is effective, but the insufficient number of gain atoms prevents lasing action. The sweet spot for squeezed lasing (blue) lies at $N_c\gtrsim 1000$ and $N_g \sim 10^2\,\text{--}\,10^3$,  where all the conditions to stabilize a clean coherent state in the squeezed basis are met. The crossover between imperfect and genuine squeezed lasing is set by condition V (see Table~\ref{tab:conditions}), which we use to mark the analytical boundary between the two regimes.
Finally, values $N_g\gtrsim 10^3$ may drive the system into non-stationary regimes or into regions where the high intracavity population can compromise the adiabatic elimination of the excited atomic levels.

Quantitatively demonstrating the presence of intracavity squeezing in the squeezed-lasing phase is subtle, since $U(1)$ symmetry enforces a steady-state solution consisting of a mixture of coherent states in the squeezed basis, summed over all possible phases~\cite{navarrete-benlloch_inducing_2014,sanchez_munoz_squeezed_2021}. While this averaged stationary solution has been shown to display genuine quantum properties (such as an improved scaling of the quantum Fisher information~\cite{sanchez_munoz_squeezed_2021}), intracavity noise reduction below vacuum is achieved only in a symmetry-broken metastable solution with a well-defined phase. A symmetry-broken solution is enforced via injection-locking, whereby one adds a small seed through an extra driving Hamiltonian $\Omega_\text{seed}(\hat a^\dagger e^{-i\phi}+\text{H.c.})$, with $\phi$ the phase we wish to enforce into the system, chosen here to minimize $\Delta \hat X$. Details on the choice of the seed are given in Methods.

In this way, we obtain a numerical counterpart of the phase diagram outlined above. \figref{fig_3}(e) shows the full-model result, computed from the steady state of the second-order cumulant expansion of \eqref{eq:me_adiabatic_full}, after breaking phase invariance with a seed field. The bivariate colormap encodes simultaneously the photon population in the squeezed basis $n_s$, and the reduction of quadrature fluctuations $\Delta \hat X$, following the 2D colorscale in the inset. The blue region marks genuine squeezed lasing---simultaneous high photon population and strong sub-vacuum intracavity squeezing---which we find confined within the regions predicted by the mean-field phase diagram in panel (a), overlaid here as dashed lines.

\subsubsection*{Spectrum of squeezing}

The presence of squeezing is also unveiled by the steady-state squeezing spectrum $S_{\hat{X}}^\text{sq}(\nu)$, plotted as a function of the sideband frequency $\nu$ measured from the reference $\omega_0$.  In the lasing phase [see Fig.~\ref{fig4}(a)], the squeezing spectrum is characterized by two main peaks, an atom-like one at $\nu\approx 0$ and a cavity-like one at $\nu\approx \Delta$, with a mirror set of peaks around $\nu=0$, since any finite frequency of $\hat a_s$ appears in the spectrum of $\hat a$ at both positive and negative frequencies. The spectrum therefore contains four peaks in total.
The atom-like peaks correspond to the actual carrier frequencies of the squeezed laser, as seen in the emission spectrum $S(\nu)$ [\figref{fig4}(b)]; this is a hallmark of superradiant lasing, with strongly suppressed cavity pulling.
These carrier peaks feature excess noise, with a linewidth set by phase diffusion and narrowed well below the decay rates [\figref{fig4}(c)]. The cavity-like peaks show reduced quadrature fluctuations, with a bandwidth $\sim\kappa+\tilde\kappa$.
Below the lasing threshold, the absence of population inversion enables cavity-atom hybridization from the coupling $g_0 \sqrt {N_g}$, which shifts the frequency of the peaks, still featuring a squeezed sideband dip due to the action of the cooling ensemble.

At very high pumping rates (close to and above $w_\text{cap}$), the squeezed cavity-like sidebands move towards $\nu=0$, while the carrier peaks move away from it. Consequently, the best output squeezing gradually fades until both lines meet at around $w^\text{p}\approx \kappa+\tilde\kappa$, where the noise is almost completely back to the vacuum level.
Due to the small $\kappa$ taken in this work, the output squeezing in the imperfect squeezed lasing phase of \figref{fig_3} is practically negligible (not shown): $\kappa+\tilde\kappa$ lies well below $w_\text{pk}$ and $w_\text{cap}$. Indeed, tuning the decay rate $\kappa$ with a fixed $w^\text{p}$, as we show in \figref{fig4}(d,e), we obtain a considerable improvement of the output squeezing, with maximum depth around $\kappa=\tilde\kappa$, in agreement with our discussion above on the spectrum of squeezed cooling [see also SI~\ref{app:emission_spec}]. This improvement comes at the expense of intracavity squeezing, but increases the lasing power of the device, since the intracavity population decreases only by a factor of $\sim2$ up to $\kappa\sim\tilde\kappa$ while $\kappa$ increases by about an order of magnitude.
These spectra correspond to the unseeded laser: over a coherence time $\sim1/\Delta\nu$ the carrier is effectively frozen relative to the drive-fixed squeezing ellipse, hence the sub-vacuum noise at $|\nu|\gg\Delta\nu$, but at longer times their relative angle diffuses, producing the anti-squeezed peak at $\nu\approx0$. Injection locking (Methods) fixes this angle and unveils the relaxation oscillations of the laser~\cite{bohnet2012a} as narrow anti-squeezed sidebands, which leave the squeezing intact. For the case shown in \figref{fig4}(e), we obtain quadrature noise of up to $\sim$\dB{5} below vacuum over a megahertz-wide band, again optimal at $\kappa\approx\tilde\kappa$ (see SI \ref{app:seed}).

\section*{Discussion}
We have proposed a mechanism enabling squeezed cooling and squeezed lasing in trapped ensembles of $^{87}$Rb atoms based on cavity-assisted Raman transitions, in which the atoms engineer the squeezed reservoir of the cavity and then lase into the squeezed cavity mode.
Our scheme generates bright squeezed light from a lasing transition, whose degree of intracavity and output quadrature-noise reduction can be selected through experimentally accessible parameters, such as the power and frequency of the transverse driving lasers.

Central to this work is the squeezed-cooling mechanism we introduce. Using the  Raman toolbox, we engineer a buffer mode that realizes an effective dissipator $\mathcal{D}[\hat a_s]$ on the squeezed cavity operator, cooling the cavity toward a squeezed steady state. %
Moreover, in the regime where the engineered decay dominates over the bare single-photon loss $\mathcal{D}[\hat a]$, the stabilization is free of the \dB{3} ceiling of intracavity parametric squeezing (e.g., in below-threshold OPOs), reaching values of up to \dB{9} in our scheme. The system can also be tuned to maximize the output squeezing instead, obtaining reductions of up to \dB{18}. Adding Raman gain on top of the squeezed cooling mechanism then yields squeezed lasing, with a bright, coherent output whose noise falls below vacuum level at sideband frequencies set by the cavity-like mode of the laser, away from the atom-like carrier due to the superradiant nature of the lasing phase, which suppresses cavity pulling.

Our work brings engineered dissipation as a new tool into the well-established setting of superradiant and Raman lasers~\cite{vrijsen2011,meiser2009,bohnet2012a, bohnet2012,meiser2010,tieri2017,debnath_lasing_2018,norcia_cold-strontium_2016,zhang_ultranarrow_2021,yu2024,zhang_extremely_2024}, where it had not previously been exploited, and opens a route to explore driven-dissipative transitions in anisotropic Dicke models realized through cavity-assisted Raman transitions~\cite{baden_realization_2014,zhiqiang_nonequilibrium_2017,zhang_dicke-model_2018,kirton_superradiant_2018,kirton_introduction_2019, stitely_nonlinear_2020, stitely_superradiant_2020,stitely_lasing_2022,ray_quantum_2022}. These directions lie within reach of current experimental platforms, whose typical parameters we have adopted throughout (Table~\ref{tab:parameters}).
Overall, this scheme provides a versatile platform for the study of squeezed lasers, directly enabling prospects of atomic metrology in which autonomous lasing and squeezing is provided within the atom cavity QED setup hosting the atoms. The scheme also offers a natural pathway toward two-mode squeezing, which could enable its use as a source of entangled light; combined with the long coherence times characteristic of laser emission, this may prove relevant for long-range quantum communication.

\bibliography{sque_laser, Atomic-Squeezed-Lasing}

\section*{Methods}
\subsection*{Atomic level scheme}
The coupling of both cooling and gain atomic ensembles to a squeezed cavity mode is implemented by directing two circularly polarized Raman lasers, transverse to the cavity axis, with frequencies $\omega_{Lr}$ and $\omega_{Ls}$, Rabi frequencies $\Omega_{r}$ and $\Omega_{s}$, and polarization $\sigma^+$ and $\sigma^-$, respectively. These lasers coherently drive transitions on the D1 line $\left(5^2S_{1/2}\leftrightarrow 5^2P_{1/2}\right)$; within our four-level system, $\Omega_s$ drives transition $|g\rangle \leftrightarrow |s\rangle$ and $\Omega_r$ drives transition $|e\rangle \leftrightarrow |r\rangle$.
 The applied weak magnetic field $\abs{B}$ is transverse to the cavity axis and aligned with its polarization direction, making $\hat{a}$ and $\hat{b}$ $\pi$-polarized. Therefore, the cavity mode $\hat a$ drives the transitions $|e\rangle \leftrightarrow |s\rangle$ and $|g\rangle \leftrightarrow |r\rangle$ (see \figref{fig_1}).

The laser frequencies are chosen to satisfy the two-photon Raman resonance condition~\cite{dimer_proposed_2007}: tuned so that a two-photon process coherently drives transitions between the hyperfine ground states $|e\rangle \leftrightarrow |g\rangle$, with the energy mismatch compensated by the emission or absorption of a cavity photon, giving rise to Hamiltonian interaction terms of the form $\hat \sigma_i \hat a^\dagger$ and $\hat \sigma_i \hat a$.
Requiring that all atoms are initially cooled and prepared so that their population lies within the effective TLS, the pure circular polarization of the laser beams prevents population leakage via other coherent two-photon processes, which are highly suppressed, as they are detuned from Raman resonance by the hyperfine ground-level splitting ($\approx6.8$ GHz).

In a similar way, to allow the cooling atomic ensemble to serve as an effective dissipative reservoir for the squeezed cavity mode, we couple it to the lossy cavity mode $\hat b$ by applying a transverse, $\sigma^+$-polarized \emph{cooling laser} with Rabi frequency $\Omega_b$, which drives transitions on the D2 line $\left(5^2S_{1/2}\to5^2P_{3/2}\right)$.
The laser frequency $\omega_{Lb}$ is selected so that, together with the lossy mode $\hat b$ of frequency $\omega_b$, it enables a resonant cavity-assisted two-photon Raman process $\ket{e}\leftrightarrow \ket{g}$, mediated by the laser-driven transition $\ket{e}\leftrightarrow \ket{u}$ followed by the cavity-induced transition $\ket{u}\leftrightarrow \ket{g}$ [see Fig.~\ref{fig_1}(b)].
Adiabatic elimination of the intermediate state $\ket{u}$ then yields an effective coupling between the two-level atomic ground-state manifold and the cavity mode, described by an interaction Hamiltonian of the form $\hat b^\dagger \hat\sigma_i$.

As detailed in SI~\ref{app:parameter_constraints_sq_mode} and \ref{app:parameter_constraints_bf_mode}, cancelling the unwanted residual terms requires Raman detunings $\Delta_r$, $\Delta_s$ and $\Delta_u$ in the GHz range. These exceed the hyperfine splittings of the D1 and D2 excited manifolds, so the lasers and the cavity also couple, at comparable detunings, to excited sublevels other than $\ket{r}$, $\ket{s}$ and $\ket{u}$~\cite{grimsmo_cavity-qed_2013}. These extra pathways leave the structure of the effective Hamiltonian unchanged. They only renormalize the effective couplings and light shifts by factors of order one, which can be compensated with the available optical knobs (laser detunings and intensities). For simplicity, we therefore keep only $\ket{r}$, $\ket{s}$ and $\ket{u}$.

Incoherent pumping of the gain ensemble is implemented with the repumping scheme proposed in Ref.~\cite{bohnet2012}, which relies on two additional transverse drives. The first is resonant with the D1 transition $\ket{F=1,m_F=0}\leftrightarrow\ket{F'=2,m_{F'}=0}$, while the second is resonant with $\ket{F=2,m_F=0}\leftrightarrow\ket{F'=1,m_{F'}=0}$. Both beams are directed in such a way that their fields contain only $\pi$ and $\sigma^-$ polarization components (see \appref{app:incoherent_pumping}), making $\ket{e}$ a dark state. In the limit where the Rabi frequency of the drives is much smaller than the natural linewidth of the D1 line, $\gamma_\text{D1}$, the excited states remain virtually unpopulated, and spontaneous emission leads to population accumulation in $\ket{e}$. Therefore, if the atoms are initially prepared in $\ket{g}$, this configuration effectively drives incoherently the $\ket{g}\rightarrow \ket{e}$ transition, resulting in the population inversion of the effective TLSs. Furthermore, the repumping scheme also mitigates, in the case of gain atoms, population leakage out of the two-level subspace via spontaneous emission from excited hyperfine sublevels virtually populated by off-resonant coupling to the different driving fields, adding robustness to the four-level system reduction against real population loss. Literature reports that this and similar configurations based on atomic repumping can achieve effective incoherent pumping rates $w^\text{p}$ of up to a few hundred kHz \cite{bohnet2012,bohnet2012a}, dominating over the rest of contributions to the transverse relaxation rate of the reduced system, like for example pure dephasing, which are hence negligible with respect to the coherence decay induced by the pumping.

The cooling atoms, however, lack a mechanism to recover the population leaked due to spontaneous emission from the virtually populated excited levels. As in previous realizations of the Dicke model with cavity-assisted Raman transitions \cite{baden_realization_2014,zhiqiang_nonequilibrium_2017,zhang_dicke-model_2018}, this limits the lifetime of the cooling atomic ensemble.

As a last remark, we note that the external laser beams considered throughout this section are directed transverse to the cavity axis, a direction along which the atoms are not confined in the Lamb-Dicke regime. Although our theoretical model does not account for atomic motional effects, these are expected to play a role in a real experimental implementation \cite{baden_realization_2014,zhang_dicke-model_2018}, requiring a renormalization of cavity couplings due to thermal motion or a treatment of inhomogeneous (e.g., Doppler) broadening of the atomic transition, which lies beyond the scope of this work.

\subsection*{Full model}
The Hamiltonian $ \hat{H} = \hat{H}_R + \hat{H}_C$ of the complete system is composed of two elements: the Raman Hamiltonian common to the two atomic ensembles, $\hat{H}_R$; and the Raman Hamiltonian driving the cooling ensemble, $\hat{H}_C$. Assuming $N_c$ atoms in the cooling ensemble and $N_g = N-N_c$ atoms in the gain ensemble, these read ($\hbar=1$),
\begin{align}
    \hat{H}_R
            &= \omega_\text{cav}\hat{a}^\dagger \hat{a}
            + \sum_{j=1}^{N} \left(\hat H_{A}^{(j)} + \hat H_{D}^{(j)} + \hat H_{I}^{(j)}\right),\\
    \hat H_{A}^{(j)} &= \omega_e \ketbra{e}{e}_j
              + \omega_r \ketbra{r}{r}_j  + \omega_s \ketbra{s}{s}_j \,,\\
    \hat H_D^{(j)} &= \frac{\Omega_r}{2} e^{it\omega_{Lr}}\ketbra{e}{r}_j
                + \frac{\Omega_s}{2} e^{it\omega_{Ls}}\ketbra{g}{s}_j
                + \text{H.c.} \,,\\
       \hat H_{I}^{(j)} &=      g_r \hat{a}^\dagger \ketbra{g}{r}_j
                + g_s \hat{a}^\dagger \ketbra{e}{s}_j  + \text{H.c};
\end{align}
and
\begin{multline}
        \hat{H}_C = \omega_b \hat{b}^\dagger \hat{b}
        + \sum_{j=1}^{N_c}
       \left\{ \omega_u \ketbra{u}{u}_j
    \right. \\
       \left.    + \left(
      g_b \hat{b}\ketbra{u}{g}_j
        + \frac{\Omega_b}{2}e^{it\omega_{Lb}}\ketbra{u}{e}_j
        + \text{H.c.}
        \right) \right\},
\end{multline}
where $g_s=2 g_r$ (see SI~\ref{app:decay_rates}). The atomic states have been indexed such that the first $N_c$ atoms belong to the cooling ensemble, and the remaining $N_g=N-N_c$ atoms, to the gain ensemble. Furthermore, without loss of generality, we assume all parameters to be real.

Moving to a rotating frame where both terms are time-independent (see \appref{app:rot_frame}), which in particular sets the frequency reference $\omega_0 = (\omega_{Lr}+\omega_{Ls})/2$ for the cavity $\hat a$, the Raman Hamiltonian becomes
\begin{subequations}
\begin{align}
         \hat{H}_R
        &= \delta_{\mathrm{cav}} \hat{a}^\dagger \hat{a}
        + \sum_{j=1}^{N} \left(\hat H_{A}^{(j)} + \hat H_{D}^{(j)} + \hat H_{I}^{(j)}\right),\\
\hat H_{A}^{(j)} &= \Delta_e \ketbra{e}{e}_j
          + \Delta_r \ketbra{r}{r}_j  + \Delta_s \ketbra{s}{s}_j \,,\\
\hat H_D^{(j)} &= \frac{\Omega_r}{2} \ketbra{e}{r}_j
            + \frac{\Omega_s}{2} \ketbra{g}{s}_j
            + \text{H.c.} \,,\\
   \hat H_{I}^{(j)} &=      g_r \hat{a}^\dagger \ketbra{g}{r}_j
            + g_s \hat{a}^\dagger \ketbra{e}{s}_j  + \text{H.c};
    \end{align}
    \label{eq:H_raman}
    \end{subequations}
and the cooling Hamiltonian results
    \begin{multline}
        \hat{H}_C = \delta_b \hat{b}^\dagger \hat{b}
        + \sum_{j=1}^{N_c}
       \left\{ \Delta_u \ketbra{u}{u}_j
    \right. \\
       \left.    + \left(
      g_b \hat{b}\ketbra{u}{g}_j
        + \frac{\Omega_b}{2}\ketbra{u}{e}_j
        + \text{H.c.}
        \right) \right\},
    \label{eq:H_cooling}
    \end{multline}
    with $\delta_\text{cav} = \omega_\text{cav}-\omega_0$, $\Delta_e = \omega_e-\frac{\omega_{Ls}-\omega_{Lr}}{2}$, $\Delta_r = \omega_r-\omega_0$, $\Delta_s = \omega_s-\omega_{Ls}$, $\delta_b = \omega_b-\omega_{Lb}-\frac{\omega_{Ls}-\omega_{Lr}}{2}$, and $\Delta_u = \omega_u-\omega_{Lb}-\frac{\omega_{Ls}-\omega_{Lr}}{2}$.

    Cavity losses and spontaneous emission processes in the atoms are modeled with the Lindblad formalism through the master equation and the standard superoperators $\mathcal{D}[\hat{O}]\equiv 2\hat{O}\hat\rho\hat O^\dagger-\{\hat O^\dagger\hat O,\hat\rho\}$ to describe incoherent channels.

    Therefore, the complete dynamics of the quantum system is described by
\begin{multline}
    \partial_t\hat{\rho}=-i\left[\hat{H},\hat{\rho}\right]+\frac{\kappa}{2}\mathcal{D}[\hat{a}]+\frac{\kappa_b}{2}\mathcal{D}[\hat{b}]+\\+\sum_{j=1}^N\sum_{\substack{k=\{r,s,u\} \\ l=\{g,e\}}}\frac{\gamma_{l,k}}{2}\mathcal{D}[\ketbra{l}{k}_j],
    \label{eq:me_comp}
\end{multline}
with $\kappa$ being the photon loss rate of mode $\hat{a}$, $\kappa_b$ the photon loss rate of mode $\hat{b}$, and $\gamma_{l,k}$ the atomic decay rate from level $\ket{k}$ to level $\ket{l}$, whose value can be expressed as a fraction of the natural linewidth of the fine-structure transition (see SI~\ref{app:decay_rates} for the specific expressions for rubidium).

\subsection*{Effective coupling to the squeezed cavity mode}
Following the arguments introduced in Ref.~\cite{dimer_proposed_2007}, we eliminate adiabatically the excited atomic levels $\ket{r}$ and $\ket{s}$, and describe the dynamics generated by $\hat H_R$ in terms of the effective TLSs formed by $\ket{g}$ and $\ket{e}$. This implies that levels $\ket{r}$ and $\ket{s}$  remain virtual, so inequalities $\abs{\Delta_{r,s}}\gg\abs{\Delta_e},\abs{\delta_\text{cav}}, \Omega_{r,s}, g_{r,s}\sqrt{N(\langle\hat a^\dagger\hat a\rangle+1)}$ must hold.  Applying the corresponding Schrieffer-Wolff transformation~\cite{qin_quantum_2024}, the effective Hamiltonian becomes
    \begin{multline}
    \hat{H}_{R,\text{eff}} = \delta \hat{a}^\dag\hat{a}
    + \sum_{j=1}^{N} \Big\{
    \delta_0 \hat{\sigma}_j^\dag \hat{\sigma}_j
    + \chi \hat{a}^\dag \hat{a} \hat{\sigma}_j^\dag \hat{\sigma}_j \\
    \quad + \left[ \left(\lambda_r \hat{a} + \lambda_{cr} \hat a^\dag \right)\hat\sigma_j^\dag + \text{H.c.}\right]
    \Big\},
    \label{eq:H_raman_eff}
    \end{multline}
with
$\delta \equiv \delta_\text{cav}-N g_{r}^2/\Delta_{r}$,
$\delta_0 \equiv \Delta_e-\tfrac{1}{4}\!\left({\Omega_{r}^2}/{\Delta_{r}}-{\Omega_{s}^2}/{\Delta_{s}}\right)$,
$\chi \equiv g_{r}^2/{\Delta_{r}}-g_{s}^2/{\Delta_{s}}$,
$\lambda_r \equiv -g_{r}\Omega_{r}/{2\Delta_{r}}$,
and $\lambda_{cr} \equiv -g_{s}\Omega_{s}/{2\Delta_{s}}$.
\paragraph*{Parameter constraints for squeezed coupling.}
The optically tunable parameters, $\omega_{Lr}$, $\omega_{Ls}$, $\Omega_r$ and $\Omega_s$ are fixed as a function of the cavity frequency $\omega_\text{cav}$ by the conditions $(\delta, \delta_0, \chi)= 0$, the equality $g_s = 2 g_r$ set by the dipole matrix elements of $^{87}$Rb transitions, and the choice of the desired $r$. In particular, we find $\omega_{Ls}\approx \omega_\text{cav} + \omega_e$, $\omega_{Lr}\approx \omega_\text{cav} - \omega_e$,  $\Omega_r = 2\cosh r\sqrt{\Delta_e \Delta_r}$ and $\Omega_s = 2 \tanh r\ \Omega_r$ (see SI~\ref{app:parameter_constraints_sq_mode} for details). These constraints also fix the single-atom coupling rate to the squeezed mode as
\begin{equation}
     g_0 = g_r \sqrt{\Delta_e/ \Delta_r},
     \label{eq:g0-simplified}
\end{equation}
which, notably, is independent from $r$.

While $\omega_\text{cav}$ and $N$ remain free parameters, they are both limited by the conditions for the adiabatic elimination of excited atomic levels. As we discuss in SI~\ref{app:parameter_constraints_sq_mode}, the cavity frequency must be close to a central value $\omega_\text{cav}=\omega_{\text{cav},0}+\delta\omega_\text{cav}$, where $\omega_{\text{cav},0}\equiv (\omega_e + 4\omega_r - \omega_s)/3\sim\omega_{D1}+2\pi\cdot 5.77\ \text{GHz}$, with the detuning from that central frequency being bounded by $\delta\omega_\text{cav}/2\pi\ll 20\ \text{MHz}$. Larger values of $\delta\omega_\text{cav}$ within this limit permit correspondingly higher values of $\Omega_r$ and $\Delta_e$, and hence of $g_0$. On the other hand, we find that $N$ must satisfy $N\ll 10^5$. Therefore, the maximum value of $N$ considered in this work is of the order of $10^4$.

\subsection*{Coupling to the buffer cavity mode}
The Raman coupling between the cooling atoms and the buffer cavity mode $\hat b$ via the excited level $|u\rangle$ is performed in a similar way. We adiabatically eliminate the state $\ket{u}$ assuming it remains virtually unpopulated, which requires $\abs{\Delta_{u}}\gg\abs{\delta_b}, \Omega_{b}, g_{b}\sqrt{N_c(\langle\hat b^\dagger \hat b\rangle+1)}$. Following the same procedure as above, we find
\begin{equation}
    \hat{H}_{C,\text{eff}}=\tilde{\delta}\hat{b}^\dagger\hat{b}+\sum_{j=1}^{N_c}\left\{ \tilde{\delta}_0\hat{\sigma}_j^\dagger\hat{\sigma}_j+\tilde{\chi}\hat{b}^\dagger\hat{b}\hat{\sigma}_j^\dagger\hat{\sigma}_j+g\left(\hat{b}\hat{\sigma}_j^\dagger+\text{H.c.}\right)\right\},
    \label{eq:H_cool_eff}
\end{equation}
with
$\tilde{\delta} \equiv \delta_b-N_cg_b^2/\Delta_{u}$,
$\tilde{\delta}_0 \equiv -\Omega_b^2/{4\Delta_{u}}$,
$\tilde{\chi} \equiv g_b^2/{\Delta_{u}}$,
and $g \equiv -\Omega_b g_b/{2\Delta_{u}}$.

\paragraph*{Parameter constraints for buffer-mode coupling.} In this case, we fix the optically tunable parameter $\omega_{Lb}$ by additionally imposing the constraint $\tilde \delta=\tilde\delta_0$. Hence, we obtain $\omega_{Lb}\approx\omega_b-\omega_e$ (see \appref{app:parameter_constraints_bf_mode}). Moreover, the conditions required to adiabatically eliminate level $\ket{u}$ restrict the magnitude of the remaining free parameters, $\Omega_b$ and $g_b$, in terms of the cavity frequency $\omega_b$. Thus, the values used in this work, listed in Table~\ref{tab:parameters}, are chosen within an experimentally realistic range to satisfy these constraints, specified in \appref{app:parameter_constraints_bf_mode}.

\subsection*{Incoherent dynamics}
\subsubsection*{Incoherent pump of gain atoms}
The ground-state TLSs of the gain atoms are incoherently pumped using two repumpers oriented such that $\ket{e}$ becomes a dark state of the system (see SI~\ref{app:incoherent_pumping} for details). First, one laser drives the population from the sublevels of the $\ket{F=1}$ manifold to those of $\ket{F=2}$ via the excited $\ket{F'=2}$ states. Then, a second pump to $\ket{F'=1}$ transfers the population to $\ket{e}$, where it accumulates in the absence of any other process.

This combination of coherent drives and spontaneous emission yields the effective Liouvillian
\begin{equation}
    \mathcal L_\text{pump}\, \hat\rho \equiv \sum_{j=N_c+1}^N \frac{w^\text p}{2}\mathcal{D}[\hat\sigma_j^\dagger],
\end{equation}
where the incoherent pumping rate $w^\text{p}$ has been shown to range from 1 to 200 kHz in experiments employing similar pumping schemes \cite{bohnet2012a,bohnet2012}.

The main limitation on the magnitude of this effective rate is the requirement that the excited-state population remain small despite repumpers being tuned to the frequency of the hyperfine transitions they drive. Consequently, their Rabi frequencies must be much smaller than the D1 linewidth $\gamma_\text{D1}$, ensuring that the atomic dynamics can be described by an effective rate-equation picture rather than by coherent dressed-state dynamics.

Based on this requirement, we establish in our numerical simulations an upper cap for $w^\text{p}$ given by $w_\text{cap}=200\ \text{kHz}$, which indicates the maximum realistic pumping rate achievable with the scheme from \appref{app:incoherent_pumping}.

\subsubsection*{Raman-induced incoherent dynamics}
Upon adiabatic elimination of the excited states under the Raman drives, spontaneous emission from those states result in effective Lindblad terms that describe local loss, gain and dephasing within the ground-state TLSs  (see \appref{app:eff_diss}). These local incoherent terms are described by a Liouvillian $\mathcal L_\text{Raman} = \mathcal L_\text{R,a} +  \mathcal L_\text{R,b}$, summing the incoherent processes induced by Raman coupling to modes $\hat a$ and $\hat b$, given respectively  by
\begin{align}
    \mathcal L_\text{R,a}\, \hat\rho &=\frac{1}{2}\sum_{j=1}^N \left\{ \gamma^\text{R,a}\mathcal{D}\left[\hat\sigma_j\right]+\gamma_\phi^\text{R,a}\mathcal{D}\left[\hat\sigma_{z,j}\right] \right.+
      \left. w^\text{R,a}\,\mathcal{D}\left[\hat\sigma_j^\dagger\right] \right\} \nonumber\\
      \mathcal L_\text{R,b}\, \hat\rho &= \frac{1}{2}\sum_{j=1}^{N_c} \left\{\gamma^\text{R,b}\mathcal{D}\left[\hat\sigma_j\right]+\gamma_\phi^\text{R,b}\mathcal{D}\left[\hat\sigma_{z,j}\right] \right\},
\end{align}
where the effective ground-state loss, dephasing, and pumping rates are, respectively,
$\gamma^\text{R,a}\equiv\gamma_{g,r}\Omega_r^2/4\Delta_r^2$,
$\gamma_\phi^\text{R,a}\equiv\gamma_{e,r}\Omega_r^2/16\Delta_r^2+\gamma_{g,s}\Omega_s^2/16\Delta_s^2$,
and $w^\text{R,a}\equiv\gamma_{e,s}\Omega_s^2/4\Delta_s^2$ for all the TLSs, and also
$\gamma^\text{R,b}\equiv\gamma_{g,u}\Omega_b^2/4\Delta_u^2$ and
$\gamma_\phi^\text{R,b}\equiv\gamma_{e,u}\Omega_b^2/16\Delta_u^2$ for the cooling TLSs.

Since these terms constitute the only source of individual loss and decoherence for ground-state TLSs, they  will be crucial to account for deviations from effective models with  purely collective dissipation, and to establish with precision the lasing threshold at the weak pumping regime.

\subsection*{Full model after elimination of the excited states}
The full model of the system after eliminating the excited states thus reads
\begin{multline}
\mathcal L_\text{full}\, \hat\rho = -i\left[\hat{H}_{R,\text{eff}}
+\hat{H}_{C,\text{eff}},\hat{\rho}\right]
    +\frac{\kappa}{2}\mathcal D[\hat{a}]
    + \frac{\kappa_b}{2}\mathcal D[\hat{b}] \\+  (\mathcal L_\text{pump}+\mathcal L_\text{R,a}+ \mathcal L_\text{R,b})\, \hat \rho .
    \label{eq:me_adiabatic_full}
\end{multline}

We will use this model to benchmark the effective descriptions via numerical simulations based on the cumulant expansion technique, implemented using \texttt{QuantumCumulants.jl} \cite{plankensteiner2022}.

\subsection*{Effective model for squeezed cooling}
To estimate the effective squeezed-cooling rates, we ignore for now the presence of gain atoms in the master equation (\ref{eq:me_adiabatic_full}), yielding
\begin{multline}
    \mathcal L_\text{cool,full}\,\hat\rho =-i\left[\hat{H}_{R,\text{eff}}^{[N_c]}
+\hat{H}_{C,\text{eff}},\hat{\rho}\right]
    +\frac{\kappa}{2}\mathcal D[\hat{a}]
    + \frac{\kappa_b}{2}\mathcal D[\hat{b}] \\+  (\mathcal L_\text{R,a}^{[N_c]}+ \mathcal L_\text{R,b})\, \hat \rho , \label{eq:me_eff_full_cooling}
\end{multline}
where $\hat{H}_{R,\text{eff}}^{[N_c]}$ and $\mathcal L_\text{R,a}^{[N_c]}$ simply denote the first $N_c$ terms in the sum over atoms in $\hat{H}_{R,\text{eff}}$ and $\mathcal L_\text{R,a}$, respectively.

We then adiabatically eliminate the buffer mode
$\hat{b}$ under the requirement that the sum of detuning and dissipation far exceeds its collective coupling rate to the ensemble (condition II in Table~\ref{tab:conditions}).
If we further disregard the dispersive coupling term in \eqref{eq:H_cool_eff}, provided the collective frequency shift it induces on $\hat b$ remains small compared to its own combined decay and detuning rates (condition IV in  Table~\ref{tab:conditions}), this yields, as shown in SI~\ref{app:eliminations_cooling}, a collective decay of the effective TLSs in the cooling ensemble $\frac{\Gamma}{2}\mathcal D[\hat S]$ (where $\hat S\equiv \sum_{j=1}^{N_c} \hat\sigma_j$), with a Purcell rate
\begin{equation}
    \Gamma = \frac{4g^2}{\kappa_b}. \label{eq:gamma_col}
\end{equation}

In a similar way, considering that the sum of the collective decay and detuning of the TLSs dominates over the collective coherent coupling to the squeezed cavity mode, and over all local incoherent processes in $\mathcal L_\text{R,a}$ and $\mathcal L_\text{R,b}$ (conditions III(i) and III(ii) in Table~\ref{tab:conditions}, respectively), the cooling ensemble itself can be adiabatically eliminated, leading to the reduced \eqref{eq:me_sq_cool} of the main text (see SI~\ref{app:eliminations_cooling}).

\subsection*{Squeezing spectrum}
To assess the squeezing of the output field, $\hat a_\text{out}=\sqrt{\kappa_\text{out}}\hat a-\hat a_\text{in}$, we compute the squeezing spectrum, which encodes the information obtained from usual photodetection measurements involving balanced homodyne detection. For the squeezed quadrature $\hat X$, assuming no additional losses ($\kappa_\text{out}=\kappa$) and perfect detection efficiency, this quantity is defined in the steady state as
\begin{equation}
    S_{\hat X}^\text{sq}(\Omega)\equiv1+\kappa\int_{-\infty}^\infty d\tau e^{i\Omega \tau}\lim_{t\to\infty}\langle:\delta\hat X(t+\tau)\delta\hat X(t):\rangle,
    \label{eq:sq_spec}
\end{equation}
where $\delta\hat X\equiv\hat X-\langle\hat X\rangle$ and $::$ denotes normal and time ordering. $S_{\hat X}^\text{sq}(\Omega)$ is normalized to the vacuum level, meaning that squeezing is present at a given sideband frequency $\Omega$ whenever its value falls below 1.

\subsection*{Non-stationary dynamics}
Within the lasing phase, the mean-field analysis predicts a region of non-stationary dynamics, bounded by the dashed black line in \figref{fig_3}(a); the honeycomb marks where the cumulant solution of the
full model is non-stationary. As derived in SI~\ref{app:stability_analysis}, this corresponds to a section of the parameter space where the nontrivial steady-state solution to the mean-field equations becomes unstable. Indeed, this type of chaotic instabilities in bad-cavity lasers were predicted as early as 1975 by Haken~\cite{haken_analogy_1975, valcarcel_semiclassical_2006}, but their relevance to Raman and superradiant lasers has only recently begun to be explored~\cite{laburthe-tolra2026}. Their role in the proposed setup and in the context of squeezed lasing will be the subject of future work.
We also note that the non-stationary region observed with the full model, while overlapping with the prediction from mean field in Fig.~\ref{fig_3}(a), occupies a smaller domain than the latter. Nonetheless, in this regime, the long-time dynamics of the system are found to range from well-defined limit cycles to fully chaotic behavior, exhibiting a rich variety of nonlinear phenomena, such as period-doubling cascades, chaotic regimes, or periodic windows. These are illustrated by the representative bifurcation diagram shown in \appref{app:stability_analysis}.

\subsection*{Phase-symmetry breaking with a seed: injection-locking}
Here, Rabi frequencies are taken as real, fixing the squeezing angle to zero and setting $\phi=\pi/2$. We cap the seed amplitude to enforce $\abs{\langle\hat a_s\rangle}/\sqrt{n_s^0}=1.2$, with $n_s^0$ being the squeezed photon population  without the seed. This criterion is enough to stabilize a well-defined phase across a considerable domain of the parameter space, while ensuring the coherent amplitude remains mainly due to lasing action rather than to the seed itself.
Note that higher seeding amplitudes would unveil squeezing for larger values of $N_g$, since the region to the right of the boundary set by condition V corresponds to a vanishing anomalous correlator $\langle \hat a_s^2\rangle$ in the absence of seed (see \appref{app:importance_sqc}). However, this would come at the expense of the coherent amplitude predominantly originating from the external field instead of from lasing, a regime that may be of interest for certain applications, but one that lies beyond our present focus on autonomous coherence buildup.

\section*{Acknowledgements}
We thank Joan Agustí and Alejandro González Tudela for useful insights and discussions. We acknowledge support by the project PID2023-149969NA-100 (SEQUOIA) funded by the Spanish Agencia Estatal de Investigación MICIU/AEI/10.13039/501100011033,  by a 2025
Leonardo Grant for Scientific Research and Cultural Creation from the BBVA Foundation,  by the project NPhoQuss funded by Programa Fundamentos 2024 from the BBVA Foundation through the grant EIC24-1-17304. R.G.d.D. acknowledges a predoctoral contract by the Spanish Agencia Estatal de Investigación with code PREP2023-001961.
\section*{Author contributions}
C.S.M. conceived and supervised the project. R.G.d.D. performed the numerical simulations. Both authors contributed to the analytical calculations, the discussion of the results, the preparation of the figures and the writing of the manuscript.
\section*{Competing interests}
The authors declare no competing interests.
\section*{Data availability}
The data supporting the findings of this study are available within the paper and its Supplementary Information. Further data are available from the corresponding author upon reasonable request.
\section*{Code availability}
The code used to generate the results of this study is available from the corresponding author upon reasonable request.

\clearpage
\section*{Extended Data}
\suppressfloats[t]

\input{table-of-parameters}
\input{table-of-conditions}

\clearpage
\section*{Supplementary Information}
\appendix

\makeatletter

\setcounter{figure}{0}
\renewcommand{\thefigure}{S\arabic{figure}}

\renewcommand{\thesection}{\Alph{section}}

\renewcommand{\theequation}{\thesection\arabic{equation}}

\@addtoreset{equation}{section}

\renewcommand{\appendixname}{}

\makeatother

\section{BEAM POLARIZATIONS AND INCOHERENT PUMPING}\label{app:incoherent_pumping}

\begin{figure*}
    \centering
    \includegraphics[width=\textwidth]{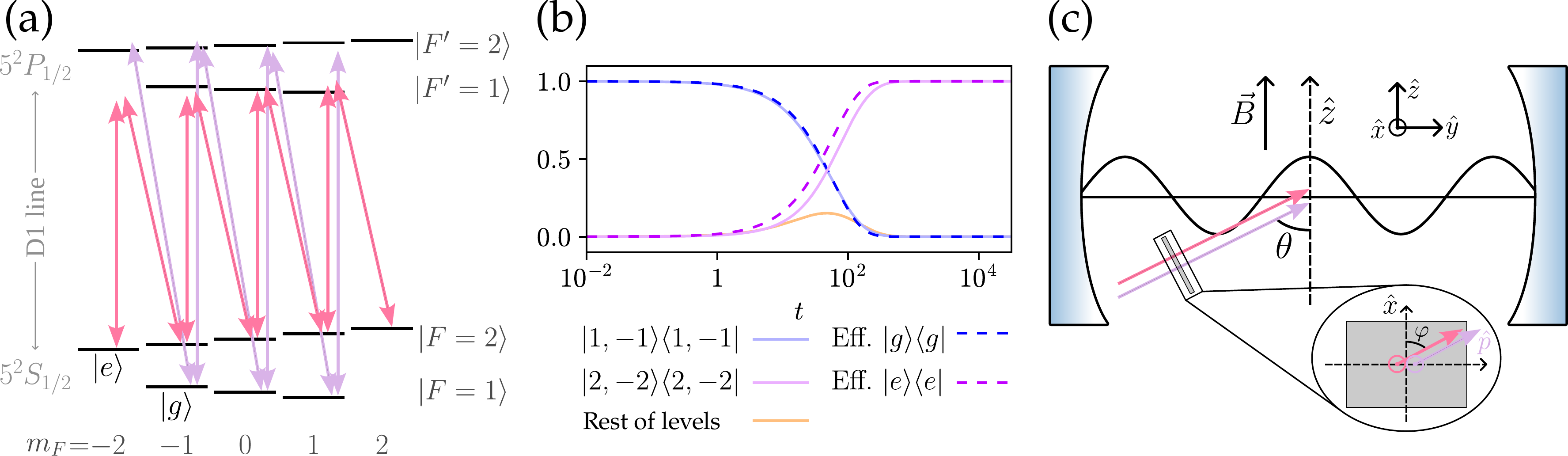}
    \caption{\textbf{Incoherent pumping scheme.} (a) Energy level diagram for the repumping drives. Both contain a combination of $\pi$ and $\sigma^-$ polarizations. (b) Time evolution of the atomic ground-state level populations under the effect of the two repumpers (solid lines). Dashed lines correspond to the master equation solution of a TLS with pumping rate $w^{p}/2\pi=3\ \text{kHz}$. In both cases, the initial state was chosen as $\ket{0}\otimes\ket{g}$. The Rabi frequencies of the repumping lasers were $0.05\cdot\gamma_\text{D1}$ for the laser driving $\ket{F=1}\leftrightarrow\ket{F'=2}$, and $0.15\cdot\gamma_\text{D1}$ for the laser driving $\ket{F=2}\leftrightarrow\ket{F'=1}$. (c) Proposed optical setup of the pumping scheme. The inset shows the polarization direction of the beams with respect to the axes of a QWP.}
    \label{fig5}
\end{figure*}

Incoherent pumping in the squeezed laser is implemented following the ideas introduced in \cite{bohnet2012,bohnet2012a}. The scheme consists of two repumping drives, each containing $\pi$ and $\sigma^-$ polarization components, which together pump the atomic population into level $\ket{e}\equiv\ket{F=2,m_F=-2}$ [see \figref{fig5}(a)]. The first drive is tuned resonant with the D1 transition $\ket{F=1,m_F=0}\leftrightarrow\ket{F'=2,m_{F'}=0}$, taking the atom out of the $\ket{F=1,m_{F}}$ manifold to $\ket{F=2,m_F}$ via spontaneous emission from the $\ket{F'=2,m_{F'}}$ states. The second drive, resonant with $\ket{F=2,m_F=0}\leftrightarrow\ket{F'=1,m_{F'}=0}$, similarly drives atoms to the $\ket{F'=1,m_{F'}}$ states, from which they spontaneously decay back into the ground manifold. With both drives applied simultaneously, repeated cycles of excitation and spontaneous decay redistribute population among the ground sublevels until it accumulates in the dark state $\ket{e}$, which is the only one that remains uncoupled from both drives. As seen in \figref{fig5}(b), this scheme effectively pumps the subspace formed by $\ket{e}$ and $\ket{g}\equiv\ket{F=1,m_F=-1}$, resembling the dynamics of a TLS with an incoherent pump.

The required polarization of the repumpers can be obtained by conveniently combining linear polarisers and quarter wave plates (QWP) in the optical setup. In our case, the magnetic field, pointing to $\hat z$, is directed perpendicularly to the cavity axis, which we will assume to be in $\hat y$. Thus, we consider a transverse beam propagating in the $yz$ plane at an angle $\theta$ with respect to the quantization axis, so that its polarization vector $\hat p$ lies in the plane spanned by $\hat x$ and $\cos\theta\hat y-\sin\theta \hat z$ [see \figref{fig5}(c)]. If the linear polarization of this beam forms an angle $\varphi$ with the fast axis of a QWP, which we can assume without loss of generality to be aligned with $\hat x$, then, the resulting polarization direction becomes $\hat p=\cos\varphi\hat x+i\sin\varphi\cos\theta\hat y-i\sin\varphi\sin\theta \hat z$. Therefore, the condition that the angles must fulfill so that the $\sigma^+$ polarization, $\hat e_+=-1/\sqrt{2}(\hat x+i\hat y)$, cancels, is $\tan\varphi=-1/\cos\theta$. If we then further impose, for example, equal $\pi$ and $\sigma^-$ components, we arrive at $\theta\approx54.7^\circ$ and $\varphi=-60^\circ$.

Furthermore, with the same derivation, we can also obtain a possible configuration for the Raman and cooling lasers, which require pure circular polarizations. In this case, to get only $\sigma^-$ component, we impose $\theta=0^\circ$ and $\varphi=-45^\circ$, and similarly, for pure $\sigma^+$, $\theta=0^\circ$ and $\varphi=45^\circ$.

\section{EXPRESSIONS FOR THE ATOMIC DECAY AND COUPLING RATES}
\label{app:decay_rates}
The atomic relaxation rates between hyperfine levels due to spontaneous emission are fractions of the fine transition lifetimes ($\gamma_\text{D1}$ and $\gamma_\text{D2}$). Following the Wigner-Eckart theorem, these are expressed in terms of the Wigner 6-$j$ symbols and the Clebsch-Gordan coefficients through its dipole matrix elements~\cite{steckRb87Dline}. In the case of rubidium, whose nuclear spin is $I=\frac{3}{2}$, we have that the relaxation rates among hyperfine sublevels within a specific line are

\begin{equation}
    \gamma^{F'm_{F'}}_{F\ m_F} = \gamma_{\text{D}i}\abs{d_{F'm_{F'},Fm_F}^{(q)}}^2
    \label{eq:gammaFF}
\end{equation}
where
\begin{equation}
    d_{F'm_{F'},Fm_F}^{(q)}
    \equiv S_{JF}^{J'F'}\braket{F,m_F|F',m_{F'}\ ;1,-q},
\end{equation}
and
\begin{equation}
    S_{JF}^{J'F'}\equiv(-1)^{F'+J+\frac{5}{2}}\sqrt{\left(2F'+1\right)\left(2J'+1\right)}\begin{Bmatrix}
    J & J' & 1 \\ F' & F & 3/2\end{Bmatrix},
\end{equation}
with $q=\{-1,0,1\}$ referring to the polarization of the Zeeman-resolved transition, and $\gamma_{\text{D}i}\ (i\in \{1,2\})$ to the natural linewidth of the relevant line (their values are provided in Table~\ref{tab:parameters}).

For clarity, we use the level symbols instead of the $(F,F',m_F,m_{F'},q)$ notation to refer to these rates throughout the text. From \eqref{eq:gammaFF} we then obtain for the D1 line
\begin{equation}
    \gamma_{g,r}=\frac{\gamma_\text{D1}}{12},\ \gamma_{e,r}=\frac{\gamma_\text{D1}}{2},\ \gamma_{g,s}=\frac{\gamma_\text{D1}}{2},\ \gamma_{e,s}=\frac{\gamma_\text{D1}}{3},
\end{equation}
and for the D2 line
\begin{equation}
    \gamma_{g,u}=\frac{5\gamma_\text{D2}}{12},\ \gamma_{e,u}=\frac{\gamma_\text{D2}}{10}.
\end{equation}

In the same fashion, the coupling strengths of cavity-induced transitions of the same line are also related through its dipole matrix elements, yielding, for the D1 line, the relation
\begin{equation}
    \frac{g_r}{g_s}=\frac{1}{2},
    \label{eq:g_rat}
\end{equation}
since $g_{r}={g_\text{cycling}}/{\sqrt{12}}$ and $g_{s}={g_\text{cycling}}/{\sqrt{3}}$, with $g_\text{cycling}$ being the reference value corresponding to the D2 cycling transition $\ket{F=2,m_{F}=2}\to\ket{F'=3,m_{F'}=3}$.

\section{ROTATING FRAME}
\label{app:rot_frame}

In order to eliminate the time dependence of the external drives in the Hamiltonian we move to the rotating frame given by the unitary transformation
\begin{equation}
    \mathcal{U}=e^{it\left[\omega_0\left(\ketbra{r}{r}+\hat a^\dagger\hat a\right)+\omega_1\ketbra{e}{e}+\omega_2\left(\ketbra{u}{u}+\hat b^\dagger\hat b\right)+\omega_3\ketbra{s}{s}\right]},
\end{equation}
with
\begin{subequations}
    \begin{eqnarray}
\omega_0&\equiv&\frac{\omega_{Ls}+\omega_{Lr}}{2},\\
 \omega_1&\equiv&\frac{\omega_{Ls}-\omega_{Lr}}{2},\\
\omega_2&\equiv&\omega_{Lb}+\frac{\omega_{Ls}-\omega_{Lr}}{2},\\
\omega_3&\equiv&\omega_{Ls}.
    \end{eqnarray}
\end{subequations}

Thus, we obtain the Hamiltonian presented in \eqref{eq:H_raman} of the Methods by applying $i\left(\partial_t\mathcal{U}\right)\mathcal{U}^\dagger+\mathcal{U}\hat H\mathcal{U}^\dagger$.

\section{PARAMETER TUNING AND CONSTRAINTS FOR THE COUPLING TO THE SQUEEZED MODE}
\label{app:parameter_constraints_sq_mode}
For $\hat H_{R,\text{eff}}$ in \eqref{eq:H_raman_eff} to describe a pure beam-splitter coupling between atoms and a squeezed cavity mode, we must tune the system parameters to achieve $\chi=\delta = \delta_0 =  0$. These conditions lead to a system of three equations which, using $g_s = 2 g_r$, read
\begin{align}
        \Delta_{s}&=4\Delta_{r}\quad (\text{sets}\, \chi=0),\label{eq:cond_r_app}\\
        \delta_\text{cav}&=\frac{Ng_{r}^2}{\Delta_{r}} \quad (\text{sets}\, \delta=0), \label{eq:cond_a_app}\\
        \Delta_e&=\frac{1}{4}\left(\frac{\Omega_{r}^2}{\Delta_{r}}-\frac{\Omega_{s}^2}{\Delta_{s}}\right) \quad (\text{sets}\, \delta_0=0).\label{eq:Delta_e_cond}
    \end{align}
    We will use these to fix the values of the free optically-tunable parameters, $\omega_{Ls}, \omega_{Lr}$, and $\Omega_r$.

\subsubsection{Raman frequencies and cavity frequency}
Solving for the two Raman laser frequencies in the first two equations (\ref{eq:cond_r_app},\ref{eq:cond_a_app}), we get
\begin{subequations}
\begin{eqnarray}
\omega_{Lr} &=& 3\omega_r-\omega_\text{cav}-\omega_s-2\xi_N, \label{eq:wlr}
\\
\omega_{Ls} &=&-2\omega_r+2\omega_\text{cav}+ \omega_s +4\xi_N, \label{eq:wls} \\
\xi_N &\equiv& \sqrt{g_r^2 N + \left(
\frac{\omega_{\text{cav}} - \omega_r}{2}
\right)^2}.
\end{eqnarray}
\label{eq:freqs_after_cond}
\end{subequations}

By defining $\epsilon \equiv g_r\sqrt N/(\omega_\text{cav}-\omega_r)$, and assuming $|\epsilon|
\ll 1$ (condition for the adiabatic elimination of excited levels), \cref{eq:wlr,eq:wls} can be approximated (taking $\omega_\text{cav}>\omega_r$) as
\begin{subequations}
\begin{eqnarray}
\omega_{Lr}
&\approx&
4\omega_r - \omega_s-2\omega_\text{cav} +2\epsilon^2(\omega_r-\omega_\text{cav}),
\label{eq:w_raman_conditions1}
\\
\omega_{Ls}
&\approx & -4\omega_r+\omega_s+4\omega_\text{cav} -4\epsilon^2(\omega_r-\omega_\text{cav}) ,
\label{eq:w_raman_conditions2}
\end{eqnarray}
\end{subequations}
which imply
\begin{eqnarray}
    \omega_0=\frac{\omega_{Lr}+\omega_{Ls}}{2} &\approx& \omega_\text{cav} - \epsilon^2(\omega_r-\omega_\text{cav}),
    \label{eq:w_raman_sum_condition}\\
    \Delta_r &\approx& (\omega_r-\omega_\text{cav})(1+\epsilon^2).
\end{eqnarray}

Furthermore, the adiabatic elimination required $|\Delta_e|\ll |\Delta_{s,r}|$. Since \cref{eq:w_raman_conditions1,eq:w_raman_conditions2}
fix the values for $\omega_{Lr}$ and $\omega_{Ls}$, the only remaining free parameter allowing to tune $\Delta_e$ is $\omega_\text{cav}$.
We must therefore choose a value of $\omega_\text{cav}$ that yields $\Delta_e \approx 0$.
We write the cavity frequency $\omega_\text{cav} = \omega_{\text{cav},0} + \delta\omega_\text{cav}$ as centered around a central value, $\omega_{\text{cav},0}$, defined so that $\Delta_e(\omega_\text{cav}=\omega_{\text{cav},0}) = 0 +\mathcal O(\epsilon^2)$. Using \cref{eq:w_raman_conditions1,eq:w_raman_conditions2}, this gives
\begin{equation}
    \omega_{\text{cav},0} = \frac{1}{3}(\omega_e + 4\omega_r - \omega_s),
    \label{eq:w_cav_condition}
\end{equation}
which has a value of $(\omega_{\text{cav},0}-\omega_\text{D1})/2\pi = 5.766\ \text{GHz}$.  Including the extra detuning $\delta\omega_\text{cav}$,  we find that
\begin{equation}
    \Delta_e \approx -3\delta\omega_\text{cav}-\epsilon^2(3\delta\omega_\text{cav}+\omega_e+\omega_r-\omega_s) .
\end{equation}

Assuming that the variation around the central frequency is very small, $\delta\omega_\text{cav}\ll \omega_e$, we then see that \cref{eq:w_raman_conditions1,eq:w_raman_conditions2} take, to zeroth order in $\epsilon$, the intuitive form of the Raman resonance conditions
\begin{subequations}
    \begin{eqnarray}
    \omega_{Ls} &\approx& \omega_\text{cav} + \omega_e + \mathcal O(\epsilon^2 ), \\
    \omega_{Lr} &\approx& \omega_\text{cav} - \omega_e + \mathcal O(\epsilon^2 ),
    \end{eqnarray}
\end{subequations}
and Raman detunings result to be
\begin{subequations}
    \begin{eqnarray}
        \Delta_r &\approx & \frac{1}{3}(\omega_s-\omega_r-\omega_e) = -2.005\cdot2\pi\ \text{GHz},\\
        \Delta_s &=&4\Delta_r\approx -8.020\cdot2\pi\ \text{GHz},
    \end{eqnarray}\label{eq:dets_raman}
\end{subequations}
so we finally have
\begin{equation}
    \Delta_e \approx -3\delta\omega_\text{cav}+3\epsilon^2\Delta_r.
    \label{eq:Deltae_simp}
\end{equation}
\subsubsection{Raman Rabi frequencies}
For a given choice of squeezing factor $r$, \eqref{eq:cond_r_app} fixes the relation between the two Raman Rabi frequencies, which is given by
\begin{equation}
\Omega_{s} = 2\Omega_{r}\tanh r.
\label{eq:cond_Omegas}
\end{equation}
Consequently, \cref{eq:Delta_e_cond,eq:cond_Omegas} set the value of $\Omega_r$ as
\begin{equation}
    \Omega_r = 2\cosh r \sqrt{\Delta_e\Delta_r}.
    \label{eq:Rabi_an}
\end{equation}

To second order in $\epsilon$ and under the assumption of $\delta\omega_\text{cav}\ll \omega_e$, using \eqref{eq:Deltae_simp}, we get
\begin{equation}
    \Omega_r = 2\sqrt 3\cosh r \sqrt{\Delta_r(\epsilon^2\Delta_r-\delta\omega_\text{cav})}.
    \label{eq:Rabi_an_simp}
\end{equation}
We also note that \eqref{eq:Rabi_an}, together with \cref{eq:cond_r_app,eq:cond_Omegas}, sets the coupling to the squeezed mode as
\begin{equation}
    g_0 = \frac{g_r \Omega_r}{2\Delta_r\cosh r}=\sqrt\frac{\Delta_e}{\Delta_r}g_r.
    \label{eq:g0-simplified-app}
\end{equation}

\subsubsection{Constraints from the adiabatic elimination}
Equation~(\ref{eq:Rabi_an_simp}) sets limits to the maximum allowed value of $\delta\omega_\text{cav}$ and $N$ through the condition $\Omega_r \ll |\Delta_r|$. Indeed, taking the square of \eqref{eq:Rabi_an_simp} we get
\begin{eqnarray}
   \epsilon^2 + \frac{\delta\omega_\text{cav}}{|\Delta_r|} \ll \frac{1}{ 12\cosh^2 r } \sim 10^{-2},
\end{eqnarray}
where we have taken $r = 1.5$, being the maximum realistic value of $r$ considered in this work.

Since
\begin{equation}
    \epsilon^2 \approx N \frac{g_r^2}{\Delta_r^2}\sim N \cdot 10^{-7},
\end{equation}
the condition $\epsilon^2 \ll 10^{-2}$ requires
\begin{equation}
    N \ll 10^5,\label{eq:lim_N}
\end{equation}
while the condition $\delta\omega_\text{cav}/|\Delta_r| \ll 10^{-2}$, given that $|\Delta_r|\approx 2\cdot2\pi\ \text{GHz}$, implies
\begin{equation}
    \delta\omega_\text{cav}/2\pi \ll 20\ \text{MHz}.\label{eq:lim_wcav}
\end{equation}

\section{PARAMETER TUNING AND CONSTRAINTS FOR THE COUPLING TO THE BUFFER MODE}
\label{app:parameter_constraints_bf_mode}
For Hamiltonian $\hat{H}_{C,\text{eff}}$ in \eqref{eq:H_cool_eff} to be resonant we set the condition $\tilde{\delta}=\tilde{\delta}_0$, which translates into
    \begin{equation}
        \delta_b =\frac{1}{\Delta_{u}}\left(N_cg_b^2-\frac{\Omega_b^2}{4}\right) \quad (\text{sets}\, \tilde{\delta}=\tilde{\delta}_0). \label{eq:cond_b_app}
    \end{equation}

\subsubsection{Raman frequency}
In this case, \eqref{eq:cond_b_app} fixes the value of $\omega_{Lb}$, the frequency of the cooling drive, yielding
    \begin{subequations}
        \begin{eqnarray}
    \omega_{Lb} &=& \frac{\omega_u+\omega_b}{2}+\frac{\omega_{Lr}-\omega_{Ls}}{2}+\zeta_{N_c},\\
        \zeta_{N_c}&\equiv&\sqrt{g_b^2N_c+\left(\frac{\omega_b-\omega_u}{2}\right)^2-\left(\frac{\Omega_b}{2}\right)^2},
        \end{eqnarray}
    \end{subequations}
and as a result
\begin{equation}
    \Delta_u = \frac{\omega_u-\omega_b}{2}-\zeta_{N_c}.
\end{equation}

Similarly to \appref{app:parameter_constraints_sq_mode}, we define $\varepsilon\equiv \sqrt{g_b^2N_c-\Omega_b^2/4}/(\omega_b-\omega_u)$, and assuming $\abs{\varepsilon}\ll1$ and $\omega_b>\omega_u$ we get
\begin{subequations}
    \begin{eqnarray}
        \Delta_u&\approx&(\omega_u-\omega_b)\left(1+\varepsilon^2\right), \label{eq:delta_u_eps}\\
        \delta_b&\approx&(\omega_u-\omega_b)\varepsilon^2. \label{eq:delta_b_eps}
    \end{eqnarray}
\end{subequations}

Recalling $\epsilon$ from \eqref{eq:freqs_after_cond} and taking \eqref{eq:delta_b_eps} to zeroth order in $\varepsilon$ and $\epsilon$, we recover the intuitive expression for the Raman resonance
\begin{equation}
    \omega_{Lb}\approx \omega_b-\omega_e+\mathcal{O}(\varepsilon^2,\epsilon^2).
\end{equation}

\subsubsection{Constraint from the adiabatic elimination}
By expressing the frequency of mode $\hat b$ with respect to $\omega_u$ as $\delta\omega_b=\omega_b-\omega_u$, condition $\abs{\varepsilon}\ll1$ implies
\begin{equation}
    N_c\ll\left(\frac{\delta\omega_b}{g_b}\right)^2+\left(\frac{\Omega_b}{2g_b}\right)^2. \label{eq:limit_Nc}
\end{equation}

The other Raman transitions had already limited the total number of atoms to $N\ll 10^5$, so by taking the maximum number of cooling atoms as $10^4$, it will be enough to satisfy either
\begin{equation}
    \frac{\delta\omega_b}{g_b}\gg10^2\quad\text{or}\quad\frac{\Omega_b}{2g_b}\gg10^2\label{eq:lims_b}
\end{equation}
to properly eliminate level $\ket{u}$. In particular, for the parameters from Table~\ref{tab:parameters} we have that $\delta\omega_b/g_b\sim10^3$.

\section{EFFECTIVE LINDBLAD DYNAMICS FROM EXCITED STATE SPONTANEOUS DECAY}
\label{app:eff_diss}
Beyond coherent dynamics, Raman lasers also induce incoherent processes in the ground-state manifold due to natural spontaneous emission into free space. Upon the adiabatic elimination of the excited states, these processes are described by a Liouvillian $\mathcal L_\text{local} = \mathcal L_\text{local,a} +\mathcal L_\text{local,b}$, composed of local terms that act into each atom individually, stemming from the Raman coupling to modes $\hat a$ (common to all atoms) and $\hat b$ (specific to cooling atoms),
\begin{align}
    \mathcal L_\text{local,a} &= \sum_{j=1}^N  \mathcal L^{(j)}_{g,r} + \mathcal L^{(j)}_{e,r} +\mathcal L^{(j)}_{g,s}+\mathcal L^{(j)}_{e,s}, \\
      \mathcal L_\text{local,b}&=\sum_{j=1}^{N_c}  \mathcal L^{(j)}_{g,u} + \mathcal L^{(j)}_{e,u},
\end{align}
where
\begin{subequations}
    \begin{align}
        \mathcal L^{(j)}_{g,r}\,\hat\rho \equiv{} &\frac{\gamma_{g,r}}{2\Delta_r^2}\,\mathcal{D}\left[\frac{\Omega_r}{2}\hat\sigma_j+g_r\hat a\hat\sigma_j\hat\sigma_j^\dagger\right] ,\label{eq:coll_ops_eff_1}\\
        \mathcal L^{(j)}_{e,r}\,\hat\rho \equiv{} &\frac{\gamma_{e,r}}{2\Delta_r^2}\,\mathcal{D}\left[\frac{\Omega_r}{2}\hat\sigma_j^\dagger\hat\sigma_j+g_r\hat a\hat\sigma_j^\dagger\right],\label{eq:coll_ops_eff_2}\\
        \mathcal L^{(j)}_{g,s}\,\hat\rho \equiv{} &\frac{\gamma_{g,s}}{2\Delta_s^2}\,\mathcal{D}\left[\frac{\Omega_s}{2}\hat\sigma_j\hat\sigma_j^\dagger+g_s\hat a\hat\sigma_j\right],\label{eq:coll_ops_eff_3}\\
        \mathcal L^{(j)}_{e,s}\,\hat\rho \equiv{} &\frac{\gamma_{e,s}}{2\Delta_s^2}\,\mathcal{D}\left[\frac{\Omega_s}{2}\hat\sigma_j^\dagger+g_s\hat a\hat\sigma_j^\dagger\hat\sigma_j\right],\label{eq:coll_ops_eff_4}\\
        \mathcal L^{(j)}_{g,u}\,\hat\rho \equiv{} &\frac{\gamma_{g,u}}{2\Delta_u^2}\,\mathcal{D}\left[\frac{\Omega_b}{2}\hat\sigma_j+g_b\hat b\hat\sigma_j\hat\sigma_j^\dagger\right],\label{eq:coll_ops_eff_5}\\
        \mathcal L^{(j)}_{e,u}\,\hat\rho \equiv{} &\frac{\gamma_{e,u}}{2\Delta_u^2}\,\mathcal{D}\left[\frac{\Omega_b}{2}\hat\sigma_j^\dagger\hat\sigma_j+g_b\hat b\hat\sigma_j^\dagger\right].\label{eq:coll_ops_eff_6}
    \end{align}
\end{subequations}

While these terms can usually be safely neglected whenever $\abs{\Delta_{r,s,u}}\gg\gamma_{D1},\gamma_{D2}$, they play an important role in our case, since they represent the only source of local decoherence that establishes the lower lasing threshold and the regimes of validity of the collective model for squeezed cooling (where local channels are disregarded).

These dissipators can be simplified, expressed just using atomic jump operators, by neglecting the terms proportional to coupling strengths in Eqs.~(\ref{eq:coll_ops_eff_1}-\ref{eq:coll_ops_eff_6}). This approximation is justified, since we can always assure that $\Omega_i\gg g_i\sqrt{\langle\hat a^\dagger\hat a\rangle}\ (i=r,s)$ and $\Omega_b\gg g_b\sqrt{\langle\hat b^\dagger\hat b\rangle}$ within the parameter constraints where the adiabatic elimination of the excited atomic states apply. This yields
\begin{subequations}
    \begin{align}
        \mathcal L^{(j)}_{g,r}\,\hat\rho \approx &\frac{1}{2} \gamma^\text{R,a}\,\mathcal{D}\left[\hat\sigma_j\right],\\
        (\mathcal L^{(j)}_{e,r} + \mathcal L^{(j)}_{g,s})\,\hat\rho \approx &\frac{1}{2}\gamma_\phi^\text{R,a}\,\mathcal{D}\left[\hat\sigma_{z,j}\right],\\
        \mathcal L^{(j)}_{e,s}\,\hat\rho \approx & \frac{1}{2}w^\text{R,a}\,\mathcal{D}\left[\hat\sigma_j^\dagger\right],\\
        \mathcal L^{(j)}_{g,u}\,\hat\rho \approx &\frac{1}{2} \gamma^\text{R,b} \,\mathcal{D}\left[\hat\sigma_j\right],\\
        \mathcal L^{(j)}_{e,u}\,\hat\rho \approx & \frac{1}{2}\gamma_\phi^\text{R,b}\,\mathcal{D}\left[\hat\sigma_{z,j}\right].
    \end{align}
\end{subequations}
where we defined
\begin{subequations}
\begin{align}
    \gamma^\text{R,a} &\equiv \frac{\gamma_{g,r}\Omega_r^2}{4\Delta_r^2}, \label{eq:gamma_ram_a}\\
    \gamma_\phi^\text{R,a} &\equiv \frac{\gamma_{e,r}\Omega_r^2}{16\Delta_r^2}+\frac{\gamma_{g,s}\Omega_s^2}{16\Delta_s^2}, \label{eq:gamma_phi_ram_a}\\
    w^\text{R,a} &\equiv \frac{\gamma_{e,s}\Omega_s^2}{4\Delta_s^2}, \label{eq:w_ram_a}\\
    \gamma^\text{R,b} & \equiv \frac{\gamma_{g,u}\Omega_b^2}{4\Delta_u^2}, \label{eq:gamma_ram_b}\\
    \gamma_\phi^\text{R,b} & \equiv \frac{\gamma_{e,u}\Omega_b^2}{16\Delta_u^2}.
    \label{eq:gamma_phi_ram_b}
\end{align}
\end{subequations}

\section{DERIVATION OF THE EFFECTIVE MASTER EQUATION FOR SQUEEZED COOLING}\label{app:eliminations_cooling}

    The master equation describing the dynamics of the cooling ensemble of $N_c$ atoms, once the excited atomic states have been adiabatically eliminated, is given by \eqref{eq:me_eff_full_cooling}. If we assume that $\hat b$ is a lossy buffer mode, i.e., $\kappa_b^2+4\tilde\delta^2\gg 4g^2 N_c \left(\langle\hat b^\dagger \hat b\rangle+1\right)$, we can eliminate this degree of freedom from the system. Additionally, in this regime, we can safely assume that $\abs{\frac{\kappa_b}{2}+i\tilde\delta}\gg\tilde{\chi}\sum^{N_c}_{j=1}\langle\hat{\sigma}_j^\dagger\hat{\sigma}_j\rangle$, because the population of the TLSs remains close to zero.

     Consequently, we can follow the Nakajima-Zwanzig formalism \cite{navarrete-benlloch2022}, ignoring the non-linear coupling term of \eqref{eq:H_cool_eff}, to eliminate $\hat{b}$. The result is that the coupling of the TLSs to the buffer mode results in an effective collective decay, yielding
        \begin{multline}
         \partial_t\hat{\rho} = -i\left[\hat{H}_{R,\text{eff}}^{N_c}+\tilde{\delta}_0\sum^{N_c}_{j=1}\hat{\sigma}_j^\dagger\hat{\sigma}_j-\frac{g^2(\tilde{\delta}-\tilde{\delta}_0)}{\frac{\kappa_b^2}{4}+(\tilde{\delta}-\tilde{\delta}_0)^2}\hat{S}^\dagger\hat{S},\hat{\rho}\right] \\
            \quad+\frac{\kappa}{2}\mathcal{D}[\hat{a}]+\frac{\Gamma}{2}\mathcal{D}[\hat{S}],
    \label{eq:me_b_elim_app}
    \end{multline}
    with $\hat{S}=\sum_{j=1}^{N_c}\hat{\sigma}_j$ being the collective lowering operator, and
    \begin{equation}
    \Gamma=\frac{g^2\kappa_b}{\frac{\kappa_b^2}{4}+(\tilde{\delta}-\tilde{\delta}_0)^2}.
\label{eq:Gamma_purcell_full}
    \end{equation}

    In order to remove the third term of the Hamiltonian in \eqref{eq:me_b_elim_app}, we fix the optically tunable parameter $\omega_{Lb}$ to meet the condition $\tilde{\delta}=\tilde{\delta}_0$ (see SI~\ref{app:parameter_constraints_bf_mode} for the specific expression), reducing the master equation to
    \begin{multline}
            \partial_t\hat{\rho} = -i\left[\tilde{\delta}_0\left(\hat{S}_z+\frac{N_c}{2}\mathbb{1}\right)+
            g_0\left(\hat a_s^\dagger \hat{S} +\hat a_s \hat S^\dagger            \right),\hat{\rho}\right]\\+\frac{\kappa}{2}\mathcal{D}[\hat{a}]+\frac{\Gamma}{2}\mathcal{D}[\hat{S}],
            \label{eq:cooling_qubits}
    \end{multline}
    where $\hat{S}_z=\frac{1}{2}\sum^{N_c}_{j=1}\hat{\sigma}^z_j$ is the collective z-spin operator.

The last step is eliminating the cooling atomic ensemble, which is possible when the collective decay of the TLSs occurs at a much faster rate than the coherent exchange of excitations with the cavity mode. To do this, we express the collective atomic operators in terms of a bosonic mode $\hat{s}$ by applying a Holstein-Primakoff transformation
    \begin{equation}
        \hat{S} = \left(\sqrt{N_c-\hat{s}^\dagger\hat{s}}\right)\hat{s},\quad\hat{S}_z=\hat{s}^\dagger\hat{s}-\frac{N_c}{2}\mathbb{1},
    \end{equation}
    which simplifies to $\hat{S}\approx\sqrt{N_c}\hat{s}$ when $N_c\gg\langle\hat{s}^\dagger\hat{s}\rangle$.

The resulting master equation reads
    \begin{equation}
        \begin{aligned}
            \partial_t\hat{\rho} =& -i\left[\tilde{\delta}_0\hat{s}^\dagger\hat{s}+\sqrt{N_c}g_0\left(\hat{a}_s^\dag\hat{s}+\hat{a}_s\hat{s}^\dag\right),\hat{\rho}\right]\\&+\frac{\kappa}{2}\mathcal{D}[\hat{a}]+\frac{N_c\Gamma}{2}\mathcal{D}[\hat{s}],
        \end{aligned}
        \label{eq:me_HP}
    \end{equation}
    and the spin bosonic mode $\hat{s}$ can be eliminated provided $(N_c\Gamma)^2+4\tilde\delta_0^2 \gg 4g_0^2 N_c\left(\langle\hat a_s^\dagger \hat a_s\rangle+1\right) $ and disregarding all local incoherent rates [\cref{eq:gamma_ram_a,eq:gamma_phi_ram_a,eq:w_ram_a,eq:gamma_ram_b,eq:gamma_phi_ram_b}], which are found to be negligible compared to the collective decay.

Finally, we end up with the following master equation
    \begin{equation}
        \partial_t\hat{\rho} = -i\left[\Delta\hat{a}_s^\dagger\hat{a}_s,\hat{\rho}\right]+\frac{\kappa}{2}\mathcal{D}[\hat{a}]+\frac{\tilde{\kappa}}{2}\mathcal{D}[\hat{a}_s],
        \label{eq:me-cooling-adiabatic}
    \end{equation}
    with
    \begin{equation}
        \Delta=-\frac{N_c\tilde{\delta}_0g_0^2}{N_c^2\frac{\Gamma^2}{4}+\tilde{\delta}_0^2}\quad \text{and}\quad \tilde{\kappa}=\frac{\Gamma g_0^2}{{\Gamma^2}/{4}+\tilde{\delta}_0^2/{N_c^2}}.
        \label{eq:kappa-tilde-full}
    \end{equation}

    Thus, under the described parameter regime, this ensemble of $N_c$ atoms enables a direct channel of single-photon decay for the squeezed cavity mode $\hat a_s$.

The full expression for the quantum fluctuations of the squeezed quadrature in the steady state of \eqref{eq:me-cooling-adiabatic} is:
   \begin{multline}
\left(\Delta\hat{X}\right)_\text{ss}^2=\frac{\left(\kappa+\tilde{\kappa}\right)^2+\left(\kappa+\tilde{\kappa}\right)\tilde{\kappa}\left(e^{-2r}-1\right)}{\left(\kappa+\tilde{\kappa}\right)^2+4\Delta^2}\\+\frac{2\Delta^2\left[\kappa\left(1+e^{-4r}\right)+2\tilde{\kappa}e^{-2r}\right]}{\left(\kappa+\tilde{\kappa}\right)^3+4\Delta^2\left(\kappa+\tilde{\kappa}\right)}.
    \label{eq:DeltaX2}
\end{multline}
which reduces to \eqref{eq:DeltaX2-simplified} when $\Delta$ is considered negligible compared to $\tilde\kappa$.

\begin{widetext}

\section{MEAN-FIELD DYNAMICS AND STABILITY ANALYSIS OF THE LASING EFFECTIVE MODEL}\label{app:stability_analysis}

The equations of motion for the expected values $\langle\hat{a}_s\rangle$, $\langle\hat{\sigma_j}\rangle$ and $\langle\hat \sigma_j^\dagger\hat{\sigma_j}\rangle$, obtained from the master equation (\ref{eq:me_lasing}), read
\begin{subequations}
    \begin{eqnarray}
\partial_t\langle\hat{a}_s\rangle&=&
-\left(\frac{\kappa+\tilde{\kappa}}{2}
+i\Delta\right)\langle\hat{a}_s\rangle
-ig_0\sum_j^{N_g}\langle\hat \sigma_j\rangle, \\
\partial_t\langle\hat{\sigma_j}\rangle&=&
-\gamma_\perp\langle\hat \sigma_j\rangle
-ig_0\langle\hat{a}_s\rangle
+2ig_0\langle\hat{a}_s\hat \sigma_j^\dagger\hat\sigma_j\rangle, \\
\partial_t\langle\hat\sigma_j^\dagger\hat{\sigma_j}\rangle&=&w-\gamma_\parallel\langle\hat\sigma_j^\dagger\hat\sigma_j\rangle-ig_0\left(\langle\hat a_s\hat\sigma_j^\dagger\rangle-\langle\hat a_s^\dagger\hat\sigma_j\rangle\right),
    \end{eqnarray}
    \label{eq:eoms-correlators}
\end{subequations}
where the longitudinal and transverse relaxation rates are $\gamma_\parallel=w+\gamma$ and $\gamma_\perp=\frac{\gamma_\parallel}{2}+2\gamma_\phi$, respectively.

We now do a mean-field ansatz, factorizing the correlators, writing the semiclassical equations as
\begin{subequations}
    \begin{eqnarray}
\partial_t \langle \hat{a}_s\rangle&=&
-\left(\frac{\kappa+\tilde{\kappa}}{2}+i\Delta\right)\langle \hat{a}_s\rangle
-ig_0 N_g\langle \hat\sigma_j\rangle, \\
\partial_t \langle \hat\sigma_j\rangle&=&
-\gamma_\perp\langle \hat\sigma_j\rangle
+ ig_0 \langle \hat{a}_s\rangle\left(2\langle \hat\sigma_j^\dagger\hat\sigma_j\rangle-1\right), \\
\partial_t \langle \hat\sigma_j^\dagger\hat\sigma_j\rangle
&=&
w -\gamma_\parallel \langle \hat\sigma_j^\dagger\hat\sigma_j\rangle + 2g_0\text{Im}\left[\langle \hat{a}_s\rangle \langle \hat\sigma_j\rangle^*\right].
    \end{eqnarray}
    \label{eq:mf_sys_app}
\end{subequations}
These are the well known Maxwell-Bloch equations for a laser.

\subsection{Normal phase}

We begin to study \eqref{eq:mf_sys_app} by considering small perturbations around the trivial or normal phase solution
\begin{equation}
    \begin{cases}
        \langle \hat{a}_s\rangle\mapsto 0+\delta\langle \hat{a}_s\rangle \\
        \langle \hat\sigma_j\rangle\mapsto 0+\delta\langle \hat\sigma_j\rangle \\
        \langle \hat\sigma_j^\dagger\hat\sigma_j\rangle\mapsto \frac{w}{\gamma_\parallel}+\delta\langle \hat\sigma_j^\dagger\hat\sigma_j\rangle
    \end{cases}
\end{equation}

Substituting and keeping just first-order terms we obtain the following linearized system
\begin{equation}
    \frac{d}{dt}\begin{pmatrix}
        \delta \langle\hat{a}_s\rangle \\
        \delta \langle\hat\sigma_j\rangle \\
    \end{pmatrix}=
    \begin{pmatrix}
        -i\Delta-\frac{\kappa+\tilde{\kappa}}{2} & -ig_0N_g  \\
        ig_0\langle\hat\sigma_z\rangle_\text{np} & -\gamma_\perp \\
    \end{pmatrix}\cdot\begin{pmatrix}
        \delta \langle\hat{a}_s\rangle \\
        \delta \langle\hat\sigma_j\rangle \\
    \end{pmatrix},
    \label{eq:lin_sys_app}
\end{equation}
with $\langle\hat\sigma_z\rangle_\text{np}\equiv\frac{w-\gamma}{\gamma_\parallel}$

The stability of this solution is then determined by the sign of the real part of the eigenvalues of the matrix in \eqref{eq:lin_sys_app},
\begin{equation}
    \lambda^\pm=-\frac{1}{4}\left(2\gamma_\perp+\kappa+\tilde{\kappa}+2i\Delta\pm\sqrt{16g_0^2N_g\langle\hat\sigma_z\rangle_\text{np}+\left(2\gamma_\perp-\kappa-\tilde{\kappa}-2i\Delta\right)^2}\right).\label{eq:eiqval_mf}
\end{equation}

Taking its real part directly leads to the following condition
\begin{equation}
    \text{Re}\left[\lambda^-\right]>0\Leftrightarrow \frac{2g_0^2N_g(w-\gamma)}{\gamma_\perp\gamma_\parallel\left(\kappa+\tilde{\kappa}\right)}>1+\left(\frac{\Delta}{\gamma_\perp+\frac{\kappa+\tilde\kappa}{2}}\right)^2,
    \label{eq:mf_thresh_app}
\end{equation}
meaning that for parameters that fulfill this inequality, the trivial solution is unstable. The system will acquire a non-zero field amplitude as a result of $U(1)$ symmetry breaking, i.e., the system enters in the lasing regime.
\subsection{Lasing phase}

To derive the expression of the lasing solution we define the photon amplitude $A\equiv \langle \hat a_s\rangle$, the atomic polarization $P = \sum_j^{N_g} \langle \hat\sigma_j\rangle$ and the magnetization $M = \sum_j^{N_g}\langle \hat\sigma_j^\dagger \sigma_j\rangle$, and solve \eqref{eq:mf_sys_app} for the steady state. However, due to the $\Delta$ term, this requires assuming that the field amplitude and the spin coherence in the long-time limit rotate with some unknown frequency. Hence, we make the ansatz $ P(t\to\infty)= P_\text{ss} e^{i\mu t}$ and $ A(t\to\infty)= A_\text{ss} e^{i\mu t}$, and solve for $A_\text{ss}$, $P_\text{ss}$ and $M$. In this way, the lasing frequency in the rotating frame is found to be
\begin{equation}
    \mu=-\frac{\gamma_\perp}{\gamma_\perp+\frac{\kappa+\tilde{\kappa}}{2}}\Delta,
\end{equation}
implying that the system oscillates at a weighted average frequency between the TLS transition frequency and the cavity frequency, which receives the name of pulling frequency ~\cite{bohnet2012}.

Knowing $\mu$, the stationary mean-field values in the lasing regime read
\begin{subequations}
    \begin{align}
     |A_\text{ss}|^2 &= \frac{N_g}{\kappa + \tilde\kappa}\left[ w-\frac{\gamma_\parallel}{2}\left(1+\frac{1}{C_{w}N_g}\right) \right]
      \label{eq:mf_A2}
     \\
     |P_\text{ss}|^{2}&=\frac{\kappa+\tilde\kappa}{2\gamma_\perp C_{w}}\abs{A_\text{ss}}^2
      \label{eq:mf_P}
     \\
      M &=\frac{N_g}{2}\left(1+\frac{1}{C_{w}N_g}\right)
       \label{eq:mf_M}
    \end{align}\label{eq:mf_complete}
\end{subequations}
where we defined the single-atom cooperativity as
\begin{equation}
    C_w \equiv \frac{\Gamma_p}{2\gamma_\perp}\cdot\frac{1}{1+\left(\frac{\Delta}{\gamma_\perp+\frac{\kappa+\tilde\kappa}{2}}\right)^2},
    \label{eq:coop}
\end{equation}
with $\Gamma_p=4g_0^2/(\kappa+\tilde\kappa)$ being the single-atom Purcell rate, and the subindex $w$ reminding it is a pump-dependent quantity. This solution confirms that the pulling frequency $\mu$ is the imaginary part of $\lambda^-$ from \eqref{eq:eiqval_mf}, after substitution of $\langle\hat\sigma_z\rangle_\text{np}$ by its mean field value in the lasing phase, which is found to be $\langle\hat\sigma_z\rangle_\text{Lp}\equiv2M/N_g-1=1/C_wN_g$.

\eqref{eq:mf_complete} is non physical unless condition (\ref{eq:mf_thresh_app}) is satisfied, which in terms of the cooperativity becomes $C_wN_g\langle\hat\sigma_z\rangle_\text{np} >1$. This condition, as described in Refs.~\cite{meiser2009,tieri2017,debnath_lasing_2018}, sets an upper and lower limit for $w$ in order to have lasing. Assuming $N_g\Gamma_p \gg \gamma,\gamma_\phi$ and $\gamma_\perp+(\kappa+\tilde\kappa)/2 \gg \abs{\Delta}$, these bounds are found to be
\begin{equation}
    w > w_\text{min} \approx \gamma\quad\text{and}\quad w < w_\text{max} \approx N_g\Gamma_p.
\end{equation}

The optimum power yielding the maximum possible photon flux can be obtained by differentiation of \eqref{eq:mf_A2}, yielding
\begin{equation}
    w_\text{pk} \approx \frac{N_g\Gamma_p}{2}, %
\end{equation}
with the corresponding optimum photon number $|A_\text{ss}|^2$ being
\begin{equation}
   n_{s,\text{pk}} =  |A_\text{ss}|^2_\text{pk} \approx \frac{N_g}{8}\left(\frac{N_g\Gamma_p}{\kappa+\tilde\kappa}\right).
\end{equation}

\subsubsection*{Chaotic instabilities}
\begin{figure}[t]
    \centering
    \includegraphics[width=0.9\textwidth]{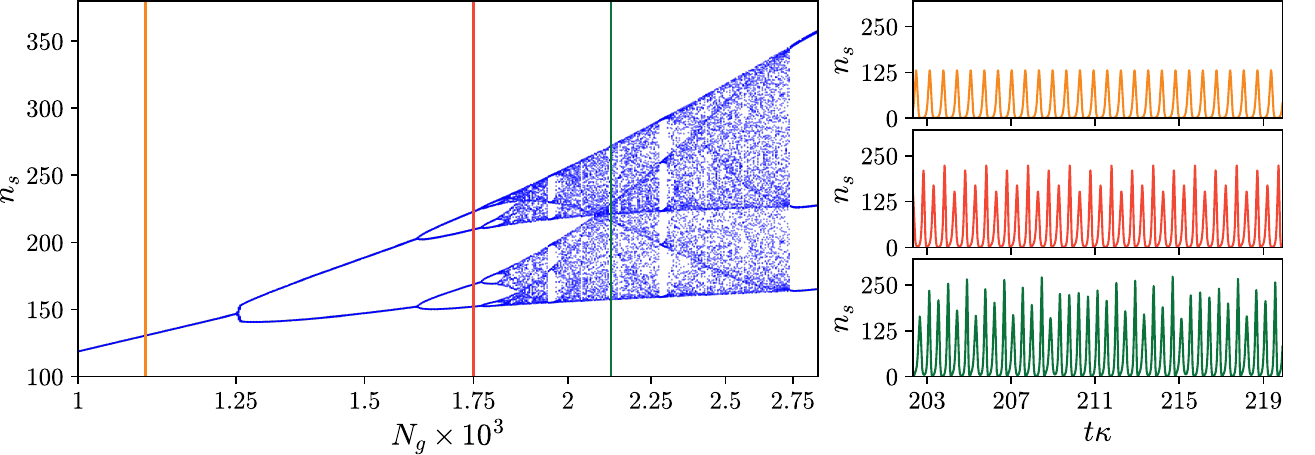}
    \caption{Bifurcation diagram calculated from the long-time dynamics of \eqref{eq:me_adiabatic_full} for $N_c=6000$, $w^\text{p}=200\ \text{kHz}$, and $r=0.5$. It shows, for a range of values of $N_g$ within the non-stationary regime, the photonic population of the squeezed mode at successive local maxima after discarding the transient dynamics. In the right panels we represent examples of the time evolution of a period-1 attractor (yellow), a period-4 attractor (red), and a strange attractor (green).}
    \label{fig6}
\end{figure}

Finally, to find the analytical stability condition of solution (\ref{eq:mf_complete}), we choose the gauge where $A$ is real and apply the following change of variables \cite{valcarcel_semiclassical_2006}
\begin{equation}
    \tilde A=\frac{A}{A_\text{ss}}\sqrt{C_wN_g\langle\hat\sigma_z\rangle_\text{np}-1},\quad \tilde P=\frac{P}{P_\text{ss}}\sqrt{C_wN_g\langle\hat\sigma_z\rangle_\text{np}-1},\quad \tilde S^z=\frac{S^z}{S^z_\text{ss}},
\end{equation}
with $S^z=2M-N_g$.

Considering again that $\gamma_\perp+(\kappa+\tilde\kappa)/2 \gg \abs{\Delta}$ and linearising the equations around (\ref{eq:mf_complete}), we end up with the following system
\begin{equation}
    \frac{d}{dt}\begin{pmatrix}
        \delta \tilde A \\
        \delta \tilde P \\
        \delta \tilde S^z \\
    \end{pmatrix}=
    \begin{pmatrix}
        -\frac{\kappa+\tilde{\kappa}}{2} & \frac{\kappa+\tilde{\kappa}}{2} & 0  \\
        \gamma_\perp & -\gamma_\perp & \gamma_\perp\sqrt{C_wN_g\langle\hat\sigma_z\rangle_\text{np}-1} \\
        -\gamma_\parallel\sqrt{C_wN_g\langle\hat\sigma_z\rangle_\text{np}-1} & -\gamma_\parallel\sqrt{C_wN_g\langle\hat\sigma_z\rangle_\text{np}-1} & -\gamma_\parallel\\
    \end{pmatrix}\cdot\begin{pmatrix}
        \delta \tilde A \\
        \delta \tilde P \\
        \delta \tilde S^z \\
    \end{pmatrix},
    \label{eq:lin_las_sys}
\end{equation}
which according to Routh–Hurwitz stability criterion will have an eigenvalue with a positive real part whenever
\begin{equation}
    \gamma_\perp+\gamma_\parallel-\frac{\kappa+\tilde\kappa}{2}<0 \quad\text{and}\quad C_wN_g\langle\hat\sigma_z\rangle_\text{np}\geq\frac{\kappa+\tilde\kappa}{2\gamma_\perp}\left(\frac{3\gamma_\perp+\gamma_\parallel+\frac{\kappa+\tilde\kappa}{2}}{
    \frac{\kappa+\tilde\kappa}{2}-\gamma_\perp-\gamma_\parallel}\right).
    \label{eq:chaos_cond}
\end{equation}

In this region of the parameter space, the dynamical system no longer evolves toward a fixed point. Instead, a Lorenz-Haken instability develops, giving rise to limit cycles or chaotic behavior \cite{haken_analogy_1975,valcarcel_semiclassical_2006, laburthe-tolra2026}. Different attractors can then be identified by constructing the bifurcation diagram of any relevant observable, as we did in \figref{fig6} for the complete model of \eqref{eq:me_adiabatic_full}. In this case, using the long-time dynamics of the photonic population in the squeezed mode, the diagram clearly reveals the onset of chaos through a period-doubling cascade, followed by the reemergence of periodic dynamics for sufficiently large atomic ensembles.

\section{CALCULATION OF THE EMISSION AND SQUEEZING SPECTRA}\label{app:emission_spec}

The output field is characterized through the calculation of relevant power spectral densities, like the emission spectrum or the spectrum of squeezing. Here we provide details on the calculation of these quantities.

\subsection{Emission spectrum}

The emission spectrum is given by the Fourier transform of the first order coherence function,
\begin{equation}
    S(\omega)=2\text{Re}\left[\lim_{t\to\infty}\int_0^\infty d\tau e^{-i\omega\tau}\langle \hat a^\dagger(t+\tau) \hat a(t)\rangle\right].
    \label{eq:spectrum_emission}
\end{equation}
Writing the two-time correlator  in terms of the squeezed cavity mode $\hat{a}_s$ requires evaluating four different two-time correlators:
\begin{multline}
    \langle \hat{a}^\dagger(t+\tau)\hat{a}(t)\rangle=\cosh^2(r)\langle \hat{a}_s^\dagger(t+\tau)\hat{a}_s(t)\rangle+\sinh^2(r)\langle \hat{a}_s(t+\tau)\hat{a}_s^\dagger(t)\rangle-\\\cosh(r)\sinh(r)\left(\langle \hat{a}_s^\dagger(t+\tau)\hat{a}_s^\dagger(t)\rangle+\langle \hat{a}_s(t+\tau)\hat{a}_s(t)\rangle\right),
\end{multline}

For this purpose, we invoke the Quantum Regression Theorem (QRT) \cite{gardiner2000quantum}, which states that, given a system of equations $\frac{d}{dt}\langle\bold{v}(t)\rangle=M\langle\bold{v}(t)\rangle$, where $\bold{v}$ is an arbitrary vector of operators, the vector $\langle \hat{O}_1(t)\bold{v}(t+\tau)\hat{O}_2(t)\rangle$ obeys the same dynamics as $\langle\bold{v}(t)\rangle$. Therefore, constructing the evolution equations for the expectation values of operators in any given master equation and homogenizing them, we can apply the QRT to the resulting system in order to compute the desired two-time correlators as
\begin{equation}
    \frac{d}{d\tau}\langle\delta\bold{v}(\tau+t)\delta \hat{a}_s(t)\rangle=M\langle\delta\bold{v}(\tau+t)\delta \hat{a}_s(t)\rangle,
    \label{eq:QRT_app}
\end{equation}
where $\langle\delta\bold{v}(t)\rangle\equiv \langle\bold{v}(t)\rangle-\bold{v}_\text{ss}$ and $\bold{v}_\text{ss}$ satisfies $M\bold{v}_\text{ss}+\bold{b}=0$.

However, (\ref{eq:QRT_app}) normally forms an infinite hierarchy of coupled equations, which we close via second-order cumulant expansion \cite{kubo1962}, expanding third-order moments as
\begin{equation}
    \langle ABC\rangle = \langle AB\rangle\langle C\rangle + \langle AC\rangle\langle B\rangle + \langle BC\rangle\langle A\rangle -2\langle A\rangle\langle B\rangle\langle C\rangle.
    \label{eq:cum_exp}
\end{equation}

As an example, we consider the full model of \eqref{eq:me_adiabatic_full} in the regime where $\tilde\kappa\gg\kappa(e^{2r}-1)$, where we can assume $U(1)$ invariance. Under this assumption, all terms carrying phase information can be ignored in the closed system after applying the cumulant expansion. Consequently, the contribution of $\langle \hat{a}_s(t+\tau)\hat{a}_s(t)\rangle$ and $\langle \hat{a}_s^\dagger(t+\tau)\hat{a}_s^\dagger(t)\rangle$ to the spectrum becomes negligible. Given that the system has also reached its steady state, $S(\omega)$ can then be completely calculated from,
\begin{subequations}
    \begin{align}
    &\frac{d}{d\tau} \mathbf C(\tau) = M \mathbf C(\tau), \\
     &\mathbf C(\tau)\equiv \begin{pmatrix}\langle \hat{a}_s^\dagger(t_\text{ss}+\tau)\hat{a}_s(t_\text{ss}) \rangle, \\ \langle \hat{b}^\dagger(t_\text{ss}+\tau)\hat{a}_s(t_\text{ss}) \rangle \\ \langle \hat{\sigma}^{c\dagger}(t_\text{ss}+\tau)\hat{a}_s(t_\text{ss}) \rangle \\ \langle \hat{\sigma}^{g\dagger}(t_\text{ss}+\tau)\hat{a}_s(t_\text{ss}) \rangle
    \end{pmatrix}\\
    &M \equiv  \begin{pmatrix}
        -\frac{\kappa}{2} & 0 & ig_0N_c & ig_0N_g \\
         0 & -\frac{\kappa_b}{2}+i\left(\tilde\delta+\tilde\chi N_c\langle\hat\sigma^{c\dagger}\hat\sigma^c\rangle_\text{ss}\right) & igN_c & 0 \\
        -ig_0\langle\hat\sigma_z^c\rangle_\text{ss} & -i\left(g\langle\hat\sigma_z^c\rangle_\text{ss}-\tilde\chi\langle\hat\sigma^{c\dagger}\hat b\rangle_\text{ss}\right) & -\gamma_\perp^c+i\tilde\chi\langle\hat b^\dagger\hat b\rangle_\text{ss} & 0  \\
        -ig_0\langle\hat\sigma_z^g\rangle_\text{ss} & 0 & 0 & -\gamma_\perp^g ,\\
    \end{pmatrix}
        \label{eq:sys_spec}
\end{align}
\end{subequations}
where $\hat\sigma^c$ and $\hat\sigma^g$ denote the lowering operators of the cooling and gain TLSs, respectively.

The solution to this system will have the form $\langle\bold{v}(\tau+t_\text{ss})\hat a_s(t_\text{ss})\rangle=e^{M\tau}\bold{v}_\text{ss}$, which we introduce in the expression for the spectrum. In this way, by diagonalizing $M$ as $M=EDE^{-1}$, we are able to formally perform the integral, yielding

\begin{multline}
    2\text{Re}\left[\lim_{t\to\infty}\int_0^\infty d\tau e^{-i\omega\tau}\langle \hat{a}_s^\dagger(t+\tau) \hat{a}_s(t)\rangle\right]=2\text{Re}\left[E\frac{1}{i\omega-D}E^{-1}\bold{v}_\text{ss}\right]_1=2\text{Re}\left[\sum_{j=1,..,4}E_{1,j}\frac{1}{i\omega-D_{j,j}}\left[E^{-1}\bold{v}_\text{ss}\right]_j\right]=\\
    =2\sum_{j=1,..,4}\Bigg\{L_j\frac{\frac{\Gamma_j}{2}}{\frac{\Gamma_j^2}{4}+(\omega_j-\omega)^2}+K_j\frac{\omega-\omega_j}{\frac{\Gamma_j^2}{4}+(\omega_j-\omega)^2}\Bigg\},
    \label{eq:spec_sum}
\end{multline}
with
\begin{align*}
    L_j = \;\text{Re}\left[E_{1,j}\left[E^{-1}\bold{v}_\text{ss}\right]_j\right],\quad
    K_j = \;\text{Im}\left[E_{1,j}\left[E^{-1}\bold{v}_\text{ss}\right]_j\right],\quad\Gamma_j =\;-2\text{Re}\left[D_{j,j}\right],\quad\omega_j = \;\text{Im}\left[D_{j,j}\right].
\end{align*}

This reveals that the contribution of correlator $\langle \hat{a}_s^\dagger(t+\tau) \hat{a}_s(t)\rangle$ to the spectrum is a weighted sum of Lorentzians and dispersive shifts whose widths $\Gamma_j$ and positions $\omega_j$ can be obtained from the eigenvalues of the matrix in \eqref{eq:sys_spec}. Additionally, the system relaxation time is associated to the smallest $\Gamma_j$, also known as Liouvillian gap. Note that this analytic calculation is just valid in the regime where engineered dissipation dominates, and thus intracavity squeezing is optimized. If we instead were in the case where $\kappa\approx \tilde\kappa$, we would need to consider the contributions from anomalous correlators too, as assuming that the system is $U(1)$ symmetric is no longer a good approximation.

Finally, we would also need to calculate the contribution from $\langle \hat{a}_s(t+\tau)\hat{a}_s^\dagger(t)\rangle$, which is easily derived from the conjugation of \eqref{eq:sys_spec}, resulting in a mirror image of (\ref{eq:spec_sum}). However, the final profile is not symmetric because each component is scaled according to the squeezing parameter $r$ that defines the Bogoliubov mode $\hat a_s$. This bimodal shape is visible in \figref{fig4}(b) from the main text, where the line narrowing due to lasing action is clearly seen too. Nonetheless, in most of the lasing regime accessible with our incoherent pumping scheme, the separation between the two peaks becomes smaller than the linewidth, so the emission resembles the monochromatic spectrum of a standard laser. Furthermore, \figref{fig4}(b) shows that the effective model of \eqref{eq:me_lasing} closely reproduces the emission properties of the system, including the position and width of the peaks, which are determined, respectively, by the imaginary and real parts of the eigenvalues $\lambda_\pm$ from \eqref{eq:eiqval_mf} substituting the mean field value $\langle\hat\sigma_z\rangle_\text{np}$ by its steady-state value obtained with the cumulant equations.

These eigenvalues define two field-atom normal modes, each matched to different Lorentzian components in the spectrum. In the normal phase, where there is no population inversion, the two modes represent hybrid atom-cavity polaritons if the collective coupling exceeds the field decay and cavity detuning. The spectrum then shows a Rabi doublet with emission bands separated by $\abs{\im{[\lambda^+-\lambda^-]}}\approx\sqrt{4g_0^2N_g(\gamma-w)/\gamma_\parallel+\Delta^2}$. In the lasing regime, emission takes place in the slow mode, $\lambda^-$, whose width is the Liouvillian gap and gives the narrow peak. The other mode keeps a width $\sim\kappa+\tilde\kappa$ and gives the broad component. For the case of Raman lasers, where $\kappa+\tilde\kappa\gg 2\gamma_\perp$, the laser mode is atom-like, meaning that coherence is mainly stored in the atomic ensemble.

\subsection{Squeezing spectrum}

The steady state squeezing spectrum, normalized so that shot noise is equal to 1, is defined in terms of the quadratures of the output field as
\begin{equation}
    S_\theta^\text{sq}(\Omega)\equiv\int_{-\infty}^\infty d\tau e^{i\Omega \tau}\lim_{t\to\infty}\langle\delta\hat X_\text{out}^\theta(t+\tau)\delta\hat X_\text{out}^\theta(t)\rangle,
    \label{eq:app_sq_spec}
\end{equation}
where $\hat X_\text{out}^\theta=\hat a_\text{out} e^{i\theta}+\hat a_\text{out}^\dagger e^{-i\theta}$ and $\delta\hat X_\text{out}^\theta\equiv\hat X_\text{out}^\theta-\langle\hat X_\text{out}^\theta\rangle$

Applying the standard input-output relation, $\hat a_\text{out}+\hat a_\text{in}=\sqrt{\kappa_\text{out}}\hat a$ and considering a vacuum input, we can rewrite the spectrum using only intracavity operators
\begin{equation}
    S_\theta^\text{sq}(\Omega)=1+\kappa_\text{out}\int_{-\infty}^\infty d\tau e^{i\Omega \tau}\lim_{t\to\infty}\langle:\delta\hat X^\theta(t+\tau)\delta\hat X^\theta(t):\rangle,
    \label{eq:app_sq_spec_2}
\end{equation}
with $::$ denoting normal and time ordering, and $\hat X^\theta=\hat a e^{i\theta}+\hat a^\dagger e^{-i\theta}$. Here we assume that all losses of mode $\hat a$ take place through the output port, so we set $\kappa_\text{out}=\kappa$.

Knowing that in the absence of external seed both the cooling and the lasing systems generate zero-amplitude steady states, i.e., $\langle\hat a\rangle=0$, if we consider the squeezed quadrature set by the Raman drives, $\hat X=\hat a+\hat a^\dagger$, and expand \eqref{eq:app_sq_spec_2} in terms of $\hat a$, we obtain
\begin{equation}
    S_{\hat X}^\text{sq}(\Omega)=1+4\kappa\lim_{t\to\infty}\text{Re}\left[\int_{0}^\infty d\tau \cos(\Omega \tau)\left(\langle\hat a(t+\tau)\hat a(t)\rangle+\langle\hat a^\dagger(t+\tau)\hat a(t)\rangle\right)\right],
\end{equation}
which using the Bogoliubov mode $\hat a_s$ becomes
\begin{multline}
    S_{\hat X}^\text{sq}(\Omega)=1+4\kappa e^{-r}\lim_{t\to\infty}\text{Re}\Bigg[\int_{0}^\infty d\tau \cos(\Omega \tau)\Big\{\cosh(r)\left(\langle\hat a_s^\dagger(t+\tau)\hat a_s(t)\rangle+\langle \hat{a}_s(t+\tau)\hat{a}_s(t)\rangle\right)\\-\sinh(r)\left(\langle\hat a_s(t+\tau)\hat a_s^\dagger(t)\rangle+\langle \hat{a}_s^\dagger(t+\tau)\hat{a}_s^\dagger(t)\rangle\right)\Big\}\Bigg].
    \label{eq:sq_spec_expand}
\end{multline}

Like in the case of the emission spectrum, we should evaluate the same four two-time correlators in order to obtain the noise distribution over the sideband frequencies. As an example, we do the analytical calculation for the effective cooling master equation (\ref{eq:me-cooling-adiabatic}), which yields the following system of equations
\begin{equation}
    \frac{d}{d\tau}\begin{pmatrix}\langle \hat{a}_s(t_\text{ss}+\tau)\hat{a}_s(t_\text{ss}) \rangle \\ \langle \hat{a}_s^\dagger(t_\text{ss}+\tau)\hat{a}_s(t_\text{ss}) \rangle
    \end{pmatrix}=
    \begin{pmatrix}
        -i\Delta-\frac{\kappa+\tilde\kappa}{2} & 0 \\
         0 & i\Delta-\frac{\kappa+\tilde\kappa}{2}
    \end{pmatrix}
    \begin{pmatrix}\langle \hat{a}_s(t_\text{ss}+\tau)\hat{a}_s(t_\text{ss}) \rangle \\ \langle \hat{a}_s^\dagger(t_\text{ss}+\tau)\hat{a}_s(t_\text{ss}) \rangle
    \end{pmatrix}.
    \label{eq:sys_cool_spec}
\end{equation}

From \eqref{eq:sys_cool_spec} we can already infer that the dips of $S_{\hat X}^\text{sq}(\Omega)$ will be centered around $\Omega=\pm\Delta$ and have a width of $\kappa+\tilde\kappa$. Furthermore, under the assumption of $\Delta$ being negligible ($ N_c \Gamma\gg |\tilde\delta_0|$), the matrix in \eqref{eq:sys_cool_spec} becomes real and the calculation of the correlators is immediate by deriving the equations of the second moments. However, instead of doing this direct computation, we can follow an input-output approach that provides a better insight of the physics in the system. The effective model is equivalent to a two sided cavity where mirror 1 (physical, with rate $\kappa$), couples to a vacuum reservoir with input $\hat a_\text{in}$, and mirror 2 (effective, with rate $\tilde \kappa$), couples to a squeezed vacuum reservoir with input $\hat a^s_\text{in}\equiv \cosh(r)\hat c_\text{in}-\sinh(r)\hat c^\dagger_\text{in}$, where $\hat c_\text{in}$ is the vacuum noise of the $\hat a_s$ channel. In the real system no squeezed light enters the cavity, as the engineered dissipation is actually simulated by the buffer mode and the cooling ensemble.

The Langevin equation for mode $\hat a$, setting $\Delta=0$, is then the one of a two-sided cavity
\begin{equation}
    \partial_t \hat a(t)=-\frac{\kappa+\tilde\kappa}{2}\hat a(t)+\sqrt{\kappa}\hat a_\text{in}(t)+\sqrt{\tilde\kappa}\hat a^s_\text{in}(t),
\end{equation}
which together with the input-output relation, $\hat a_\text{out}=\sqrt\kappa\hat a-\hat a_\text{in}$, in the Fourier space, yields
\begin{equation}
    \hat X_\text{out}(\Omega)=\rho(\Omega)\hat X_\text{in}(\Omega)+t(\Omega)\hat X^s_\text{in}(\Omega),
    \label{eq:output_quad}
\end{equation}
where $\hat X_\text{in}$ and $\hat X^s_\text{in}$ are the quadratures of the vacuum and squeezed-vacuum inputs, respectively, $\rho(\Omega)=\frac{\kappa}{\frac{\kappa+\tilde\kappa}{2}-i\Omega}-1$, and $t(\Omega)=\frac{\sqrt{\kappa\tilde\kappa}}{\frac{\kappa+\tilde\kappa}{2}-i\Omega}$.

In \eqref{eq:output_quad}, $\rho(\Omega)$ is the reflection amplitude of the vacuum arriving at mirror 1, while $t(\Omega)$ is the amplitude of the squeezed field transmitted from the input of mirror 2 to the output of  mirror 1. This transmission component is the one carrying the squeezing of the output field, and both amplitudes satisfy the relation $\abs{t}^2+\abs{\rho}^2=1$.

The most favorable situation occurs for $\kappa=\tilde\kappa$ at $\Omega=0$. In this case, $\abs{t(0)}=1$ and $\abs{\rho(0)}=0$, so the cavity is transparent for the squeezed field arriving at mirror 2 and negative interference cancels the amplitude of the reflected vacuum field. We should then expect the output squeezing to be optimal. The opposite happens in the regime where $\tilde\kappa\gg\kappa$, for which $\abs{t(0)}\to 0$ and $\abs{\rho(0)}\to 1$. Squeezed noise does not transmit through mirror 1, which only reflects vacuum noise, even though the intracavity squeezing is high if condition ${\tilde{\kappa}}/{\kappa}\gg\left(e^{2r}-1\right)$ is satisfied.

To demonstrate these last statements we can compute the squeezing spectrum from \eqref{eq:app_sq_spec} in the Fourier space
\begin{equation}
    2\pi\delta(\Omega+\Omega')S^\text{sq}_{\hat X}(\Omega)=\langle\hat X_\text{out}(\Omega)\hat X_\text{out}(\Omega')\rangle,
\end{equation}
which gives
\begin{equation}
    S_{\hat X}^\text{sq}(\Omega)=\abs{\rho(\Omega)}^2+\abs{t(\Omega)}^2e^{-2r}=1-\frac{4\kappa\tilde\kappa}{(\kappa+\tilde\kappa)^2+4\Omega^2} \left(1-e^{-2r}\right)
    \label{eq:sq_spec_cool}
\end{equation}
when using $\langle\hat X_\text{in}(\Omega)\hat X_\text{in}(\Omega')\rangle=2\pi\delta(\Omega+\Omega')$, $\langle\hat X^s_\text{in}(\Omega)\hat X^s_\text{in}(\Omega')\rangle=2\pi\delta(\Omega+\Omega')e^{-2r}$, and the fact that both inputs are uncorrelated.

From \eqref{eq:sq_spec_cool}, it is very clear that the minimum of the spectrum is at $\Omega=0$, and that the output squeezing is maximized when $\kappa=\tilde\kappa$, for which $S_{\hat X}^\text{sq}(0)=e^{-2r}$. Using \eqref{eq:DeltaX2} with $\Delta=0$ we note that in this regime of perfect output coupling, the noise reduction of the intracavity field is reduced to just $\Delta\hat X^2=(e^{-2r}+1)/2$, which coincides with the fundamental \dB{3} limit of OPOs for $r\to\infty$. On the other hand, in a favorable regime for intracavity squeezing where ${\tilde{\kappa}}/{\kappa}\gg\left(e^{2r}-1\right)$, $S_{\hat X}^\text{sq}(0)\to 1$, indicating that most of the squeezing is lost through the engineered dissipation channel. For example, using $\kappa$ and $\tilde\kappa$ from  Table~\ref{tab:parameters} and $r=0.5$, we obtain from \eqref{eq:sq_spec_cool}, $S_{\hat X}^\text{sq}(0)=\dB{-0.98}$, and from \eqref{eq:DeltaX2-simplified}, $\Delta\hat X=\dB{-3.74}$. If we instead increase $\kappa$ so that it matches the value of $\tilde\kappa$ we get $S_{\hat X}^\text{sq}(0)=\dB{-4.34}$ and $\Delta\hat X=\dB{-1.65}$.

\end{widetext}

\section{COOLING TLSs EXCITATION DUE TO LASING ACTION}\label{app:failure}

When combining the gain and the cooling ensemble within the same system, as the photonic population of mode $\hat a_s$ increases, cooling atoms can eventually acquire a non-negligible population and cease to act as the lossy channel described in \appref{app:eliminations_cooling}. To illustrate this, we analyze a simplified scenario, considering a collection of TLSs with collective decay subjected to a strong coherent field, which represents the intracavity field built during lasing action. If we assume a large field amplitude $\alpha$, the classical component dominates over quantum fluctuations, so the dynamics are well described by equation
\begin{equation}
    \partial_t\hat\rho=-i\left[\tilde\delta_0\hat S_z+g_0\alpha(\hat S^\dagger+\hat S),\hat\rho\right]+\frac{\Gamma}{2}\mathcal{D}[\hat S].
    \label{eq:qubit_coll}
\end{equation}

Taking the expectation values of the normalized spin operators $s_i\equiv \frac{2\langle\hat S_i\rangle}{N_c},\ i\in\{x,y,z\}$, mean-field equations read
\begin{subequations}
\begin{eqnarray}
    \partial_t s_x&=&-\tilde\delta_0s_y+\frac{N_c\Gamma}{2}s_z s_x, \\
    \partial_t s_y&=&\tilde\delta_0s_x+\frac{N_c\Gamma}{2}s_z s_y-2g_0\alpha s_z, \\
    \partial_t s_z &=&2g_0\alpha s_y-\frac{N_c\Gamma}{2}\left(s_x^2+s_y^2\right).
    \label{eq:mf_qubit_coll}
\end{eqnarray}
\end{subequations}

Solving \eqref{eq:mf_qubit_coll} in the long-time limit, and using spin conservation, $s_x^2+s_y^2+s_z^2=1$, we obtain that the steady-state population of each TLS is
\begin{equation}
    \langle\hat\sigma^\dagger\hat\sigma\rangle=\frac{1}{2}\left(1-\sqrt{\mathcal{S}+\sqrt{\mathcal{S}^2+\frac{4\tilde\delta_0^2}{(N_c\Gamma)^2}}}\right)
    \label{eq:pop_cooltls}
\end{equation}
with $\mathcal{S}=\frac{1}{2}-\frac{8(g_0\alpha)^2+2\tilde\delta_0^2}{(N_c\Gamma)^2}$.

In order for the cooling TLSs to remain in the ground state, we impose $\langle\hat\sigma^\dagger\hat\sigma\rangle\ll 1$ in \eqref{eq:pop_cooltls}. Doing so, we obtain the following inequality for the system parameters
\begin{equation}
    4\alpha^2g_0^2 \ll (N_c\Gamma)^2+4\tilde\delta_0^2.
    \label{eq:cond_cool_TLS}
\end{equation}

Thus, for the squeezed cooling scheme not to break down by lasing action, the intracavity population $\alpha^2$ must fulfill condition (\ref{eq:cond_cool_TLS}) .

\begin{figure}
    \centering
    \includegraphics[width=1\linewidth]{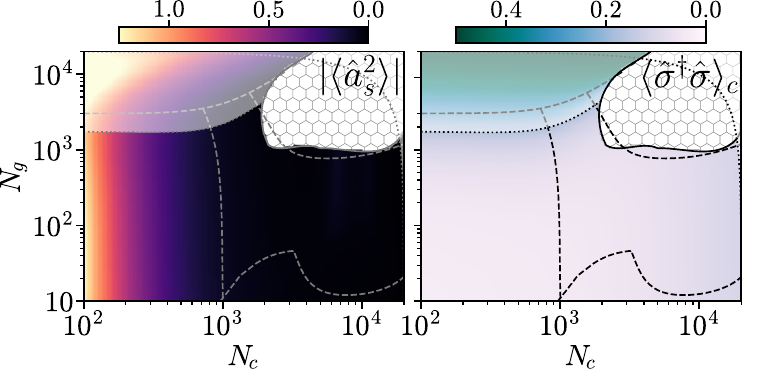}
    \caption{(a) Absolute value of the quadratic correlator $\langle\hat a_s^2\rangle$ and (b) population of the TLSs from the cooling ensemble, $\langle\hat\sigma^\dagger\hat\sigma\rangle_c$. Both were calculated from \eqref{eq:me_adiabatic_full}. Results are shown across the $(N_c,N_g)$ space for $w^\text{p}=\min(w_\text{pk},w_\text{cap})$ and $r=0.5$, the same choice as for \figref{fig_3}(d) in the main text.}
    \label{fig7}
\end{figure}

In the main text, we have colored in red the regions of the phase diagrams where \eqref{eq:cond_cool_TLS} is violated. As we can see in \figref{fig7}, these regimes exhibit partial excitation of the cooling ensemble, together with undesired squeezing correlations in the squeezed basis, i.e., nonzero $\langle\hat a_s^2\rangle$. Both are signs that the squeezed cooling mechanism fails to engineer the desired dissipation into the squeezed mode.

\section{IMPORTANCE OF SQUEEZED COOLING}\label{app:importance_sqc}

\begin{figure}[b]
    \centering
    \includegraphics[width=0.9\linewidth]{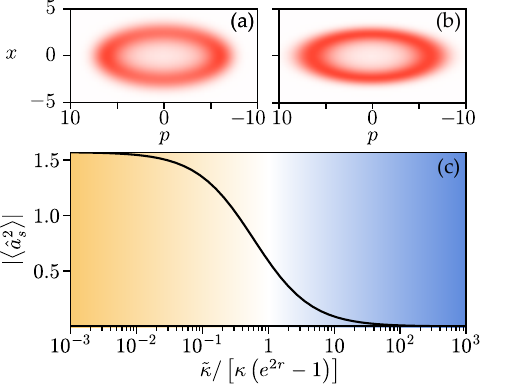}
    \caption{(a) Wigner function of the photonic steady state of \eqref{eq:me_simple_app} in the bare basis and in the lasing regime. (b) Same object as in (a) but substituting $\mathcal{D}[\hat{a}]$ for $\mathcal{D}[\hat{a}_s]$ in the master equation. (c) Change of the steady-state value of $\abs{\langle \hat a_s^2\rangle}$. The sum $\tilde{\kappa}+\kappa$ is fixed and we progressively change the relative weight of the rates. Parameters used were $w/2\pi\approx48\ \text{kHz}$, $\tilde{\kappa}+\kappa = w/40$, $C_w=1.5$ and $r=0.5$.}
    \label{fig8}
\end{figure}

In this section, we assess the consequences of neglecting the mechanism of squeezed cooling for squeezed lasing.

We consider the simplified master equation
\begin{equation}
    \begin{split}
        \partial_t\hat{\rho} =& -ig_0\left[\hat{a}_s^\dag\hat{\sigma} +\text{H.c.},\hat{\rho}\right]+\frac{w}{2}\mathcal{D}[\hat{\sigma}^\dagger]+\frac{\kappa}{2}\mathcal{D}[\hat{a}],
    \end{split}
    \label{eq:me_simple_app}
\end{equation}
which describes a standard one-atom laser for the Bogoliubov mode $\hat{a}_s$ with bare single-photon loss.

In the squeezed basis, the decay channel behaves as a coupling of the cavity to a squeezed reservoir \cite{navarrete-benlloch_inducing_2014,scully_zubairy_1997}, so instead of reaching a phase-diffused mixture of coherent states in the long-time limit, the system will evolve toward a mixture of displaced squeezed vacua
\begin{multline}
    \text{tr}_\text{TLS}[\hat{\rho}]\equiv\varrho=\frac{1}{2\pi}\int_0^{2\pi}\ketbra{\Psi}{\Psi}d\phi\\\text{with}\quad\ket{\Psi}=\hat{D}_s\left(\sqrt{n_s}e^{i\phi}\right)\hat{S}_s(-r)\ket{0}_s, \label{eq:ss_sq_basis}
\end{multline}
where $\ket{0}_s$ is the vacuum state in the squeezed basis, $\hat{S}_s(r)=e^{\frac{r}{2}\left(\hat{a}_s^2-\hat{a}_s^{\dagger2}\right)}$, and $\hat{D}_s(\alpha)=e^{\alpha\hat{a}_s^\dagger-\alpha^*\hat{a}_s}$.

Expressing state (\ref{eq:ss_sq_basis}) in the bare basis we obtain
\begin{multline}
    \varrho=\frac{1}{2\pi}\int_0^{2\pi}\ketbra{\Psi}{\Psi}d\phi\quad\text{with} \\ \ket{\Psi}=\hat{D}\left(\sqrt{n_s}\cosh(r)e^{i\phi}-\sqrt{n_s}\sinh(r)e^{-i\phi}\right)\ket{0},
    \label{eq:ss_bare_basis}
\end{multline}
which is a mixture of coherent states.

This simplified derivation shows the need of engineering effective squeezed dissipative environments like the one presented in \appref{app:eliminations_cooling} in order to observe states with quadrature fluctuations below shot noise limit in Dicke systems.

As an example, we show in \figref{fig8}(a) the Wigner function of a steady state of \eqref{eq:me_simple_app} in the lasing phase. Though its shape resembles that of squeezed lasing states [\figref{fig8}(b)], the calculation of $\langle\hat a_s^2\rangle$ reveals it is not $U(1)$-symmetric in the squeezed basis. When introducing dissipation into the squeezed mode, the system undergoes a crossover as its rate $\tilde\kappa$ increases [\figref{fig8}(c)], passing from an imperfect to a genuine squeezed lasing regime. In the limit $\tilde{\kappa}\gg\kappa\left(e^{2r}-1\right)$, we find $\langle\hat a_s^2\rangle\to0$, consistent with the steady state approaching a phase-diffused mixture of coherent states in that basis.

\section{SEEDED SQUEEZING SPECTRUM}\label{app:seed}

\begin{figure}
    \centering
    \includegraphics[width=0.9\linewidth]{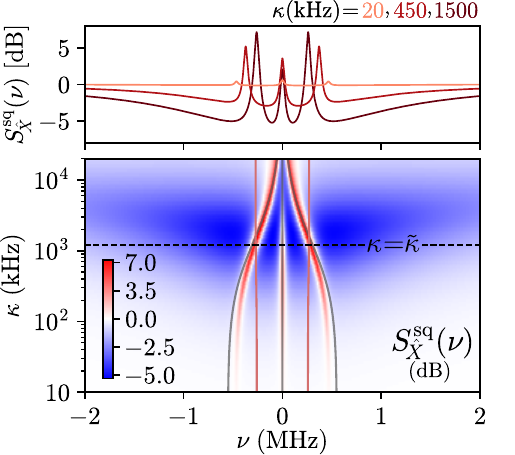}
    \caption{Steady-state squeezing spectrum of the $\hat X$ quadrature as a function of $\kappa$. Results, expressed in decibels, correspond to \eqref{eq:me_adiabatic_full} with an added seed $i\Omega_\text{seed}(\hat a -\hat a^\dagger)$, where $\Omega_\text{seed}/2\pi=1.6\ \text{kHz}$. The rest of parameters are the same as in \figref{fig4}(e). The upper panel shows the spectrum profile at selected values of $\kappa$. The positions of the peaks and dips obtained for the effective model of \eqref{eq:me_lasing} are marked as grey (excess noise) and red (reduced noise) solid curves.}
    \label{fig9}
\end{figure}

Adding an external seed in the lasing system gives a non-zero value to $|\langle\hat a_s\rangle|$, revealing a pair of excess-noise peaks on both sides of the carrier (see \figref{fig9}) which were hidden in the unseeded second order cumulant spectrum from SI \ref{app:emission_spec}. We identify these as relaxation oscillations of the laser.
Since the seed phase $\phi=\pi/2$ aligns the carrier with the squeezed quadrature $\hat X$, the fluctuations of $\hat X$ are, to linear order, amplitude fluctuations of a bright field, so that $S_{\hat X}^\text{sq}(\nu)$ also gives the shot-noise-normalized relative intensity noise (RIN) of the output: the relaxation-oscillation peaks are the usual RIN peaks of a laser, while the dips below unity correspond to sub-shot-noise intensity fluctuations.

Relaxation oscillations correspond to the damped exchange of energy between the field and the stored inversion as the system evolves towards its stable fixed point, continuously re-excited by quantum noise. When the eigenvalues associated with these peaks acquire a positive real part, the oscillations become undamped. The stable fixed point undergoes a Hopf bifurcation and the system enters the non-stationary regime observed in \figref{fig_3} for the unseeded model.

Despite the presence of this new feature with excess noise, if the seed has the appropriate phase (see Methods), the central peak is considerably reduced with respect to the unseeded case, resulting, according to the relation
\begin{equation}
    \left(\Delta\hat X\right)^2=1+\int^\infty_{-\infty}d\Omega\frac{S_{\hat X}^\text{sq}(\Omega)-1}{2\pi\kappa},
\end{equation}
in subvacuum intracavity quadrature noise, which is optimal in the regime $\tilde\kappa\gg\kappa(e^{2r}-1)$.

\end{document}

%% file: table-of-parameters.tex
\begin{table}[t]
\begin{ruledtabular}
\begin{tabular}{lllc}
Param. &  Description & $\text{Value}/2\pi $ & Ref. \\
\hline
\multicolumn{4}{c}{\textbf{Natural parameters}} \\
\hline
$\omega_e $ & Hyperfine splitting      & $6.828\  \mathrm{GHz}$  & \cite{steckRb87Dline} \\
$g_r $ & Cav. coupling to $\ket{g}\leftrightarrow \ket{r}$    & $0.78\ \mathrm{MHz}$  &  \cite{arnold_self-organization_2012}\\
$g_s $ & Cav. coupling to $\ket{e}\leftrightarrow \ket{s}$    & $1.56\ \mathrm{MHz}$  &  \cite{arnold_self-organization_2012}\\
$g_b $ & Cav. coupling to $\ket{g}\leftrightarrow \ket{u}$    & $3\ \mathrm{MHz}$  &  (\ref{eq:lims_b})\\
$\kappa $ & Decay rate of lasing mode   & $70\ \mathrm{kHz}$  &  \cite{arnold_self-organization_2012} \\
$\kappa_b $ & Decay rate of buffer mode   & $20\ \mathrm{MHz}$  &  \\
$\gamma_\text{D1} $ & Natural linewidth of D1 line   & $5.75\ \mathrm{MHz}$  & \cite{steckRb87Dline}  \\
$\gamma_\text{D2} $ & Natural linewidth of D2 line   & $6.06\ \mathrm{MHz}$  & \cite{steckRb87Dline}  \\
$\omega_\text{D1} $ & Central freq. of D1 transition & $377.107\ \mathrm{THz}$ & \cite{steckRb87Dline} \\
$\omega_\text{D2} $ & Central freq. of D2 transition & $384.230\ \mathrm{THz}$ & \cite{steckRb87Dline} \\
$\omega_r$ & Freq. of $\ket{g}\leftrightarrow\ket{r}$ transition & $\omega_\text{D1}+3.76\ \mathrm{GHz}$ & \cite{steckRb87Dline} \\
$\omega_s$ & Freq. of  $\ket{g}\leftrightarrow\ket{s}$ transition & $\omega_\text{D1}+ 4.57\ \mathrm{GHz}$ & \cite{steckRb87Dline} 
\\
$\omega_u$ & Freq. of  $\ket{g}\leftrightarrow\ket{u}$ transition & $\omega_\text{D2}+ 4.04\ \mathrm{GHz}$ & \cite{steckRb87Dline} \\
$\omega_{\text{cav},0} $ & Central lasing-cavity freq. & $\omega_\text{D1}+5.77 \ \mathrm{GHz}$ & (\ref{eq:w_cav_condition})\\
$\delta\omega_\text{cav} $ & Cav. detuning ($\omega_\text{cav}-\omega_{\text{cav},0}$) & $1 \ \mathrm{MHz}$ & (\ref{eq:lim_wcav}) \\
$\omega_b$ & Freq. of cavity mode $\hat b$ & $\omega_u+2\ \text{GHz}$ & (\ref{eq:lims_b}) \\
$\Delta_r
 $ & Raman detuning $\ket{e}\leftrightarrow\ket{r}$ & $-2\ \mathrm{GHz}$ & (\ref{eq:dets_raman}) \\
$\Delta_s $ & Raman detuning $\ket{g}\leftrightarrow\ket{s}$ & $ -8\ \mathrm{GHz}$ & (\ref{eq:dets_raman}) \\
$\Delta_u $ & Raman detuning $\ket{e}\leftrightarrow\ket{u}$ & $-2\ \mathrm{GHz}$ & (\ref{eq:delta_u_eps}) \\
$\Delta_e $ & TLS splitting (rot. frame) & $ -6.09 \ \mathrm{MHz}$ & (\ref{eq:Deltae_simp}) \\
$\Omega_r $ & Rabi freq. of $\ket{e}\leftrightarrow\ket{r}$ drive & $0.22 \cdot\cosh r \ \mathrm{GHz}$ & (\ref{eq:Rabi_an})\\
$\Omega_s $ & Rabi freq. of $\ket{g}\leftrightarrow\ket{s}$ drive & $0.44 \cdot\sinh r \ \mathrm{GHz}$ & (\ref{eq:cond_Omegas})\\
$\Omega_b $ & Rabi freq. of cooling drive & $242\ \text{MHz}$ & (\ref{eq:lims_b}) \\
$\abs{B}$ & External magnetic field & $2.7\ \text{G}$ & \cite{bohnet2012} \\
\hline 
\multicolumn{4}{c}{\textbf{Effective parameters}} \\
\hline 
$g $ & Buffer mode coupling rate & $181\ \text{kHz}$ & (\ref{eq:H_cool_eff})\\
$\tilde\delta_0 $ & Detuning of cooling TLSs &  $7.3\ \text{MHz}$ & (\ref{eq:H_cool_eff})\\
$\tilde\chi$ & Buffer mode NL coupling rate & $-4.5\ \text{kHz}$ & (\ref{eq:H_cool_eff})\\
$\Gamma $ & Cooling-TLS Purc. rate (via $\hat b$) & $6.5\ \text{kHz}$ & (\ref{eq:Gamma_purcell_full}) \\
$\Gamma_p $ & Gain-TLS Purc. rate (via $\hat a_s$) & $9.2\ \text{kHz}$ & (\ref{eq:purcell-p}) \\
$g_0 $ & Squeezed mode coupling rate & $43\ \text{kHz}$ & (\ref{eq:g0-simplified-app})\\
$\tilde\kappa $ & Decay rate of squeezed mode & $0.73\ \text{MHz}$ & (\ref{eq:kappa-tilde-full})\\
$\Delta $ & Squeezed mode detuning & $-0.27\ \text{MHz}$ & (\ref{eq:kappa-tilde-full}) \\
$w^\text{R,a} $ & Pumping induced by $\Omega_s$ & $1.45\cdot\sinh^2 r\ \text{kHz}$ & (\ref{eq:w_ram_a}) \\
$\gamma^\text{R,a} $ & Decay induced by $\Omega_r$ & $1.45\cdot\cosh^2 r\ \text{kHz}$ & (\ref{eq:gamma_ram_a}) \\
$\gamma^\text{R,b} $ & Decay induced by $\Omega_b$ & $9.22\ \text{kHz}$ & (\ref{eq:gamma_ram_b}) \\
$\gamma_\phi^\text{R,a} $ & Dephasing induced by $\Omega_r$ & $(2.17\cdot\cosh^2 r+$ & (\ref{eq:gamma_phi_ram_a}) \\
&  and $\Omega_s$ & $0.54\cdot\sinh^2 r)\ \text{kHz}$ &  \\
$\gamma_\phi^\text{R,b} $ & Dephasing induced by $\Omega_b$ & $0.55\ \text{kHz}$ & (\ref{eq:gamma_phi_ram_b}) \\
$w^\text{p} $ & Pumping induced by repumpers & $(0.16-2800)\ \text{kHz}$ & \cite{bohnet2012} \\

\end{tabular}
\end{ruledtabular}
\caption{Summary of the system parameters used in this work. The effective values were calculated using $N_c=3000$ and $N_g=400$, which, for the range of $w^\text{p}$ considered, correspond to the squeezed lasing and squeezed vacuum regime. Specifically, it is the same choice as for the cross sections in \figref{fig_3}.}
\label{tab:parameters}

\end{table}

%% file: table-of-conditions.tex
\begin{table*}[t]
\begin{ruledtabular}
\begin{tabular}{ll}
\textbf{Condition} &  \textbf{Expressions} \\
\hline
I. Elimination of excited atomic states $\ket r, \ket s,$ and $\ket u$& (i) $\abs{\Delta_r}\gg \abs{\Delta_e},\abs{\delta_\text{cav}},\ \Omega_r,\ g_r\sqrt{N}\sqrt{\langle\hat a^\dagger \hat a\rangle+1},\ \gamma_\text{D1}$ \\
 & (ii) $\abs{\Delta_s}\gg \abs{\Delta_e},\abs{\delta_\text{cav}},\ \Omega_s,\ g_s\sqrt{N}\sqrt{\langle\hat a^\dagger \hat a\rangle+1},\ \gamma_\text{D1}$  \\
 & (iii) $\abs{\Delta_u}\gg \abs{\delta_b},\ \Omega_b,\ g_b\sqrt{N_c}\sqrt{\langle\hat b^\dagger \hat b\rangle+1},\ \gamma_\text{D2}$ \\
 & (iv) $w^\text{p}\leq w^\text{cap}$ \\
 \hline
II. Elimination of mode $\hat b$ & $\kappa_b^2+4\tilde\delta^2\gg 4g^2 N_c \left(\langle\hat b^\dagger \hat b\rangle+1\right)$ \\
\hline
III. Elimination of the cooling ensemble & (i) $(N_c\Gamma)^2+4\tilde\delta_0^2 \gg 4g_0^2 N_c\left(\langle\hat a_s^\dagger \hat a_s\rangle+1\right) $  \\
& (ii) $\sqrt{(N_c\Gamma)^2+4\tilde\delta_0^2} \gg \ \gamma^\text{R,a},\ \gamma_\phi^\text{R,a},\ w^\text{R,a},\ \gamma^\text{R,b},\ \gamma_\phi^\text{R,b}$ \\
& (iii) $N_c\gg\langle\hat{s}^\dagger\hat{s}\rangle$ \\

\hline
IV. Disregard of the nonlinear coupling term $\propto \tilde \chi$ in \eqref{eq:H_cool_eff} & $\abs{\frac{\kappa_b}{2}+i\tilde\delta}\gg\tilde{\chi}\sum^{N_c}_{j=1}\langle\hat{\sigma}_j^\dagger\hat{\sigma}_j\rangle$ \\
\hline
V. Negligible effect of the natural cavity decay on intracavity squeezing & ${\tilde{\kappa}}/{\kappa}\gg\left(e^{2r}-1\right)$  \\
\end{tabular}
\end{ruledtabular}
\caption{Summary of parameter conditions mentioned along the text. Conditions I-IV establish the regime at which the system described by \eqref{eq:me_comp} effectively behaves like the reduced model in \eqref{eq:me_lasing}, while V defines the requirement for bare single-photon loss not affecting noise reduction of the intracavity field.}
\label{tab:conditions}

\end{table*}